\documentclass{aa}
\newcommand{\mss}{mag arcsec$^{-2}$}
\newcommand{\Bmo}{\rm ${\mu _B}$(0) }

\newcommand{\kms}{km s$^{-1}$}

\newcommand{\Msol}{M$_\odot$}
\newcommand{\Lsol}{L$_{\odot,B}$}

\newcommand{\obc}{[OBC97]}
\newcommand{\aeq}{$^<\!\!\!_\sim$}
\newcommand{\am}[2]{$#1'\,\hspace{-1.7mm}.\hspace{.0mm}#2$}
\newcommand{\degree}{$^\circ$}
\newcommand{\HI}{\mbox{H\,{\sc i}}}
\newcommand{\nan}{Nan\c{c}ay}

\input{epsf}
\usepackage{graphicx}
\usepackage{rotating}
\begin{document}
\title{A new  H{\large \bf I} catalog of Low Surface Brightness galaxies out to z=0.1}
\subtitle{Tripling the number of massive LSB galaxies Known}

\author{K. O'Neil\inst{1}\thanks{Work done while at Arecibo Observatory}
\and G. Bothun\inst{2}
\and W. van Driel\inst{3} \and D. Monnier Ragaigne\inst{3}}
\institute{NRAO, PO Box 2, Green Bank WV 24944, U.S.A.
\and University of Oregon, Physics Department, Eugene OR, 97402, U.S.A.
\and Observatoire de Paris, Section de Meudon GEPI, CNRS UMR 8111 and Universit\'e de Paris 7, 5 place Jules Janssen, 92195 Meudon Cedex, France}
\offprints{K. O'Neil, \email{koneil@nrao.edu}}
\date{Received 14 January 2004 / Accepted }

\abstract{
Using both the Arecibo 305m and the Nan\c{c}ay decimetric 100-m class radio telescopes,
we have observed the \HI\ line of 116 Low Surface Brightness (LSB)
galaxies from the Bothun et al. 1985 subset of 
LSB galaxies in the Uppsala General Catalog.  The observations had a detection rate of 70\%, resulting in the new determination of \HI\
properties for 81 galaxies.  Surprisingly, roughly half of the detected objects (38)
have M$_{HI} \ge 10^{10}$ \Msol, placing them into the category of massive
LSB galaxies.  As previously only $\sim$18 of these `Malin 1 cousins' were
known, our results have more than tripled the number of these fascinating and enigmatic
systems known.

Combining our results with previous studies done on the Bothun et al. catalog
results in a well-defined catalog of \HI\
properties of 526 LSB galaxies ranging in redshift space from $0 \le z \le 0.1$.
With this catalog in hand, we have been able to explore the parameter space
occupied by LSB galaxies more completely than has been previously possible.
In agreement with previous studies, our results show LSB
galaxies with some of the most extreme properties of disk galaxies, including
M$_{HI}$/L$_B$ ratios often exceeding 10 \Msol/L$_{\odot,B}$.  

}

\titlerunning{Tripling the Number of Massive LSBGs}
\authorrunning{O'Neil et al.}
\maketitle

\section{Introduction}

Low Surface Brightness (LSB) galaxies, those objects with central surface brightness
at least one magnitude fainter than the night sky, are now well established
as a real class of galaxies with properties distinct from those that define
the Hubble Sequence (e.g. Impey \& Bothun \cite{impey97}; Bothun et al. \cite{bothun97}
and references
therein). Yet considerable uncertainty still exists as to both the range in properties
of these galaxies and their number density in the z $\le$ 0.1 Universe.
As LSB galaxies encompass many of the 'extremes' in galaxy properties,
gaining a firm understanding of LSB galaxy properties and number counts is vital
for testing galaxy formation and evolution theories.
Additionally, in this era of precision
cosmology (e.g. Spergel et al. \cite{spergel03}) it becomes increasingly important
to determine the relative amounts of baryons that are contained in
galaxy potentials compared to those that may comprise the Intergalactic Medium.
As long as the possibility exists that optical and baryonic luminosity functions
are significantly biased toward galaxies with relatively high surface brightness,
we will not be able to obtain a clear understanding of the percentage of baryons 
which lie in galaxies.

The `traditional' perception of LSB galaxies is 
that they are low mass, fairly blue systems
with relatively high M$_{HI}$/L$_B$ values and low metallicities
(Bergmann et al. \cite{bergmann03}; Bell et al. \cite{bell99};
Gerritsen \& de Blok \cite{gerritsen99}; de Blok et al. \cite{deblok95}).  
Consequently, LSB galaxies are often equated with young dwarf galaxies
(Dekel \& Woo \cite{dekel03}; Cabanela \& Dickey \cite{cabanela02};
Jimenez et al. \cite{jimenez98}).  However, the simple, consistent, and repeatable
observation that LSB disk galaxies can be found at any value of circular
velocity (or total mass)
 is often under-appreciated, reinforcing the
erroneous perception that LSB galaxies are strictly low
mass systems.  The current set of observations for all LSB galaxies shows
them to have a remarkably diverse array of properties:

\begin{itemize}
\item Very red LSB galaxies have been found by optical surveys searching in both 
the B and V optical bands (Burkholder et al. \cite{burkholder01}; O'Neil et al. \cite{oneil97});  
\item Recent \HI\ surveys by, e.g. Burkholder et al. (\cite{burkholder01}) and 
O'Neil et al. (\cite{oneil00b})
show no tendency toward LSB systems having lower than average \HI\ masses
and/or total dynamical masses.  The LSB sources with \HI\ detections in
Burkholder et al. (\cite{burkholder01}), for example,
have on average $\langle log(M_{HI}/M_\odot)\rangle$ = 9.0 $\pm$ 0.6, compared to 
$\langle log(M_{HI}/M_\odot)\rangle$ = 9.4 $\pm$ 0.6 for that survey's
High Surface Brightness (HSB) galaxy sample -- both samples have the same mean and range of values
of circular velocity;
\item A number of massive (M$_{HI} \ge 10^{10}$\Msol)
LSB systems have been found, including Malin 1 -- 
the largest disk galaxy known to date (Matthews et al. \cite{matthews01}; Pickering et al. \cite{pickering97}; 
Davies et al. \cite{davies88}; Sprayberry et al. \cite{sprayberry95});
\item Although the average LSB galaxy does have lower metallicities than the average HSB
galaxy, LSB galaxies with near-solar abundances have been found
(Bell \cite{bell00});
\item While many LSB systems with extremely high M$_{HI}$/L$_B$ values have been found,
the argument that $\langle M_{HI}/L_B\rangle_{LSB} > \langle M_{HI}/L_B\rangle_{HSB}$ may be
due more to the nature of the \HI\ LSB galaxy surveys -- since redshifts for LSB systems
are often found by searching in \HI, LSB galaxies without detectable quantities of \HI\
do not make it into databases, artificially raising the value of
$\langle M_{HI}/L_B\rangle_{LSB}$.  Countering this trend, the
recent survey  of Burkholder et al.
(\cite{burkholder01}) obtained optical redshifts for a large sample of LSB galaxies which were 
also observed in \HI.  The resultant catalog contains LSB systems with M$_{HI}$/L$_B$
as low as 10$^{-6}$ \Msol/\Lsol.
\end{itemize}

While none of the above results contradict the idea that the {\it average} LSB galaxy
is less evolved than the {\it average} HSB galaxy, they do show that we have not yet come
close to fully sampling the LSB galaxy parameter space.  
In addition, it should be emphasized that there may still be large numbers of 
LSB galaxies with properties beyond our present detection limits (e.g. Sabatini et al. \cite{sabatini03}).

One issue related to the study of LSB galaxies lies in their
potential contribution 
to the total baryonic content of the Universe.  A wide variety of data and opinions have
been offered on this topic.   Blind \HI\ surveys, such as those done by Kilborn et al. (\cite{kilborn02}),
Rosenberg \& Schneider (\cite{rosenberg00}), Kraan-Korteweg et al. (\cite{kraan99}), Knezek (\cite{knezek99})
and Zwaan et al. (\cite{zwaan97}) detect no massive LSB galaxies in the 
local (z \aeq\ 0.025) Universe, leading to the conclusion  that LSB galaxies could not
be significant contributors to the local \HI\ mass function (HIMF) and subsequently to the
baryonic mass function.  
Similar conclusions have also been reached based on optical surveys (e.g. Blanton et al. \cite{blanton03};
Cross et al. \cite{cross01}).
Such a conclusion, however, is valid only if
one believes that a fair and representative volume of the Universe has been
probed in both the optical and 21-cm bands, and that the properties
of the objects found in these surveys is well understood.

The analysis of the WMAP results have provided a precision measure of
the total baryonic content of the Universe, albeit in a model dependent way.
In normalized units, the contribution of baryons to the total energy density
of the Universe is 4.5\%.  This contribution is 12 times higher than that
which is contained in optically selected catalogs of galaxies which do not
contain LSB galaxies (Shull \cite{shull03}; Bothun \cite{bothun03})
which either means that a) we are missing most
of the galaxy population in our current catalogs, or b) 95\% of the baryonic
content of the Universe is in the (presumably warm) IGM.
At present, the needed baryons in the warm IGM (alternative b) remain undetected.   
Moreover, the best fitting
6-parameter WMAP model contains a curious result -- the global ratio of
baryonic energy density to dark matter  energy density is 0.19. This
is two times larger than the baryonic mass fraction that is typically
found in disk galaxies from analysis of rotation curves and clearly implies
the existence of undetected baryons.   It is possible that the undetected population
is LSB galaxies.

It has been repeatedly shown that the question of how much a galaxy type contributes to the
baryon (and matter) density in the Universe is twofold --
what is the number density of massive galaxies of that type, and what is the slope
of the low mass luminosity/baryon density function.  Whereas numerous surveys are (and have been)
undertaken to address the latter question (e.g. Sabatini et al. \cite{sabatini03},
Mobasher et al. \cite{mobasher03}), here we aim to look
into the former.  In a nutshell, we would like to examine the questions ``How many
massive LSB disk galaxies are in the z $\le$ 0.1 Universe, and what are their
properties?''

To shed light on these questions, we have re-investigated one of the first
surveys designed to measure the space density of LSB galaxies. 
Bothun et al. (\cite{bothun85}) measured H I masses and redshifts for a well-defined sample of LSB
galaxies  chosen from the UGC (Nilson \cite{nilson73}). While that survey
was reasonably successful, there were a significant number of 21-cm non-detections which
indicated the presence of either a gas poor population of LSB disks or
large LSB disks beyond the redshift sensitivity limit of the 
survey (about 12,000 \kms).  The primary impetus behind  our re-investigation, then,
was the upgrade of the 305m Arecibo telescope. 
With its improved sensitivity,
baselines, and spectrometer, the Arecibo telescope is able to cover a much larger region
of velocity space with higher sensitivity than has previously been achievable at
21-cm.  Consequently, we chose re-examine Bothun et al.'s original \HI\ catalog,
compiling all objects with published \HI\ properties (both those detected in the 
original catalogs and those which were not) and re-observing the objects which
were originally not detected.  The end result is a well-defined catalog of \HI\
properties of 526 LSB galaxies ranging in redshift space from $0\le z < 0.1$.
With this catalog in hand, we are more fully able to explore the parameter space 
occupied by LSB galaxies.

\section{Catalog selection}
The objects chosen for observation were taken from the sample of LSB 
galaxies selected from the UGC by Bothun et al. (\cite{bothun85}), which  
meet the following criteria:
\begin{itemize}
\item Morphological type (as given in UGC) of Sb or later;
\item Blue surface brightness of $\rm \langle \mu\rangle = m_{pg} + 5\;log(D) + 8.89 - 0.26 > 25.0$ \mss\
(where m$_{pg}$ is the photographic magnitude from the UGC, $D$ is the diameter in arcmin, 8.89 is the
conversion from arcmin to arcsec, and 0.26 is an average conversion from m$_{pg}$ to m$_B$.)
This definition, used by Bothun et al. (\cite{bothun85}), is typically equivalent to the more commonly
used classification of a galaxy as LSB if it has $\mu_{B,0}^{observed} \ge $23.0 mag arcsec$^{-2}$,
although the uncertainty of this equation, combined with the high error of the UGC magnitudes
($\pm$0.5) means this catalog may contain objects with central surface brightnesses as high as
$\mu_{B,0}$ = 22 \mss.
\item Declination in the range 0$^\circ$--36$^\circ$.
\end{itemize}

A total of 1865 galaxies meet the above criteria.  Of these, 571 were randomly selected
and observed by Bothun et al. (\cite{bothun85}) with 334 being detected (and later confirmed) in \HI.
In the intervening years, 107 of the galaxies not originally detected by Bothun et al.
have had their  \HI\ properties published in the literature. The remaining 130 objects
break down as follows:
\begin{itemize}
\item 4 galaxies were rejected from the catalog due to being part of an interacting group 
(UGC~3737, UGC~11564, UGC~11027, and UGC~11057);
\item 15 of the objects listed as detections by Bothun et al. (\cite{bothun85}) were not confirmed
by follow-up observations and are considered non-detections for this paper;
\item 86 galaxies from the original list were observed for this paper;
\item The remaining 25 galaxies, all of which were listed as non-detections by
Bothun et al. (\cite{bothun85}) were not observed for this paper (due to scheduling restrictions),
and have had no other follow-up \HI\ observations.
\end{itemize}

To compensate for the `missing' galaxies, an additional 24 galaxies, randomly chosen from the
original catalog of 1865 objects, were observed.  Additionally, 6 LSB galaxies
from the catalog of O'Neil et al. (\cite{oneil97}) which fit into the scheduled observing time
on the Arecibo telescope were also observed.  The final result is a catalog
of 561 \HI\ observations of LSB galaxies, 116 of which are new
observations.  A complete list of the 116 newly observed galaxies and their optical 
properties is given in Table~\ref{tab:props} \& \ref{tab:obs}.  
The \HI\ and optical properties of all other galaxies, both those originally
detected by Bothun et al. (\cite{bothun85}) and those later detected in the literature, can be found
on-line at http://www.gb.nrao.edu/$\sim$koneil/data.

\section{Observations}

For efficiency, observations were made using both the
Arecibo 305m and the 100-m class Nan\c{c}ay radio telescopes.
Details of observations with the individual telescopes are given below,  published properties
of the galaxies are given in Table~\ref{tab:props}, and a complete
listing of the observations is given in Table~\ref{tab:obs}.

\subsection{Arecibo Observations}

Observations with the Arecibo 305m telescope were taken between 16 May, 1999 -- 02 April, 2002.
Data was taken using the L-wide and L-narrow receivers and two separate polarization channels. 
Two different correlator set-ups were
used -- one for galaxies with published velocities and one for galaxies whose velocity was unpublished
at the time of the observations.  The objects without published velocity were observed using
four 50MHz band passes with 2048 lags and 3-level sampling, giving an unsmoothed
resolution of 5.2 km/s at 1420 MHz.  The frequency range covered for these observations
was either 1270 -- 1425MHz or 1238 -- 1423MHz.  (Two different frequency ranges were used
due to the increasingly frequent presence of a strong radar at $\sim$1240 -- 1260 MHz.)
In both cases sufficient overlap was given to the individual correlator boards 
to effectively eliminate any problems which otherwise may have 
arisen due to poor performance in the outer 50 channels of each correlator.

If an observed galaxy already had a published velocity, only two correlator boards were used.
In this case each board was centered at the galaxy's redshifted \HI\ line, recorded a 
separate polarization channel, and had a total of 2048 lags across its 12.5 MHz band.  
Here, all the data had 9-level sampling.

All Arecibo data was observed in position-switched mode, with a minimum of one
10 minute on- and off-source observation pair taken of each galaxy, followed by a 20 second ON/OFF 
observation of a calibrated noise diode.  In most cases a galaxy was observed at least
twice using this method, and often as many as 4--6 times.
System temperatures were obtained from the noise diode observations, using the
predetermined temperature values from Arecibo (available on-line at http://www.naic.edu).
A gain curve for the telescope was obtained through reducing all observations taken of
standard continuum calibrators, by any project, during the observing period.
A complete description of the procedures used to determine the gain curve in this manner can be
found in Heiles (\cite{heiles01}).  Additionally, observations were made of standard continuum
calibrators every 2-3 hours during the project observations, with the results checked
against the determined telescope gain, to insure no anomalous behavior occurred in the
hardware during observations.  As a result we can confidently state that the
calibration corrections are good to within 10\% (and often much better).
A thorough discussion of the errors involved can be found
in O'Neil (\cite{oneil04}).  
 
\begin{figure*}
\centering
\includegraphics[width=6.0in]{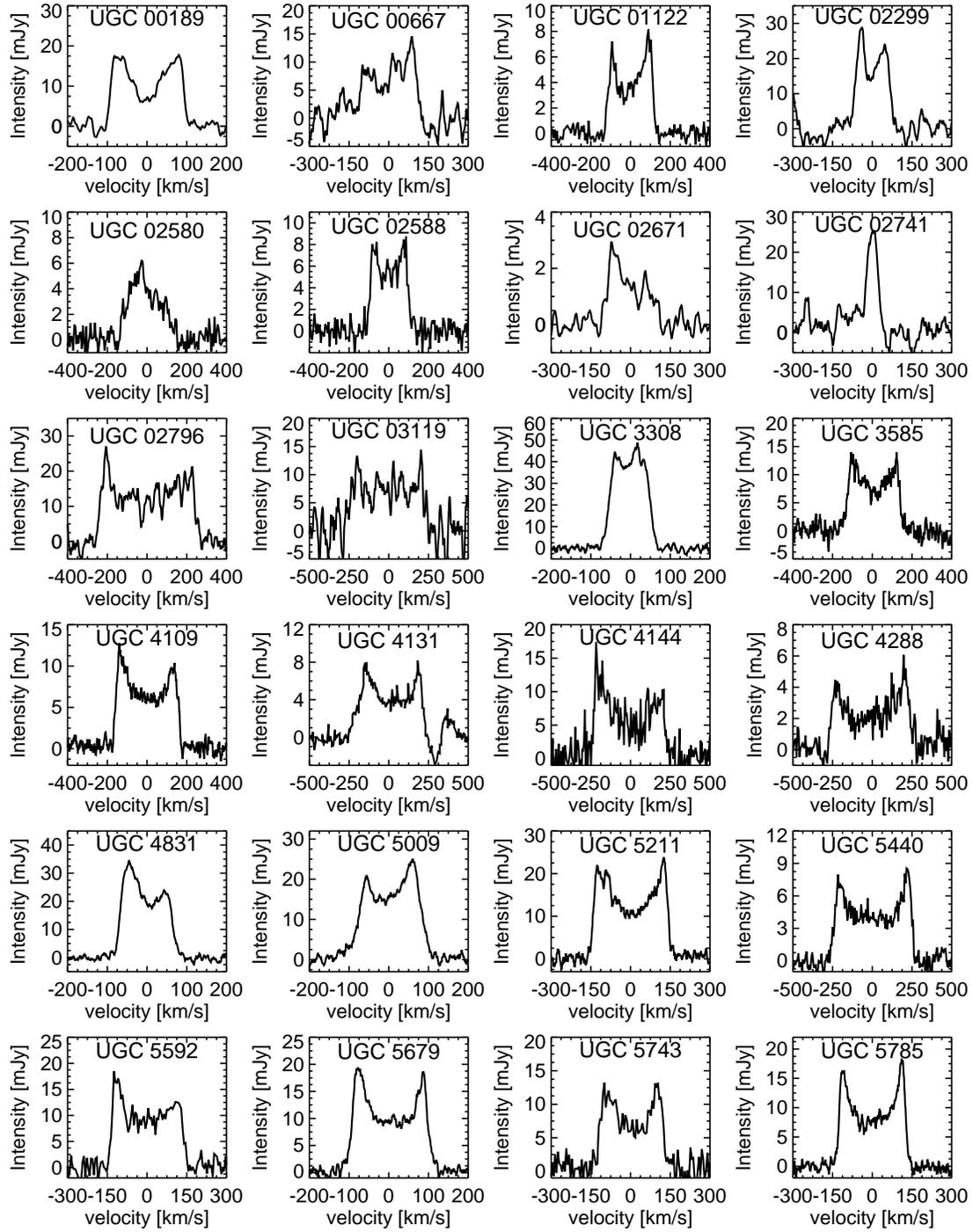}
\caption{\HI\ spectra of all detected galaxies, excluding cases of possible confusion.
Channel resolution is given in Table~\ref{tab:obs}. The velocity scale is heliocentric 
and according to the conventional optical definition.}
\label{fig:hi_spectra}
\end{figure*}

\addtocounter{figure}{-1}
\begin{figure*}
\centering
\includegraphics[width=6.0in]{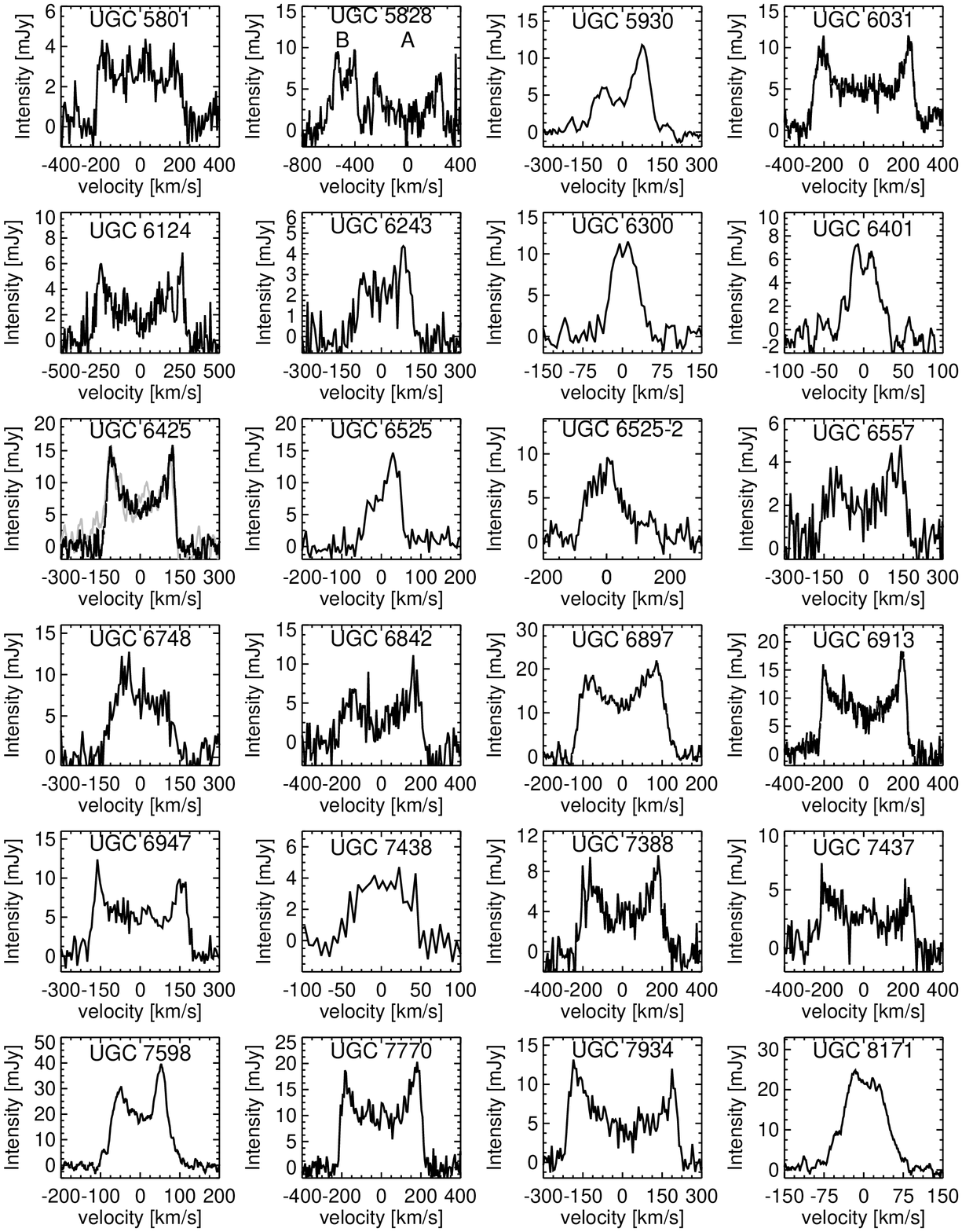}
\caption{{\it cont.} \HI\ spectra of all detected galaxies, excluding cases of possible confusion.
Channel resolution is given in Table~\ref{tab:obs}.}
\end{figure*}

\addtocounter{figure}{-1}
\begin{figure*}
\centering
\includegraphics[width=6.0in]{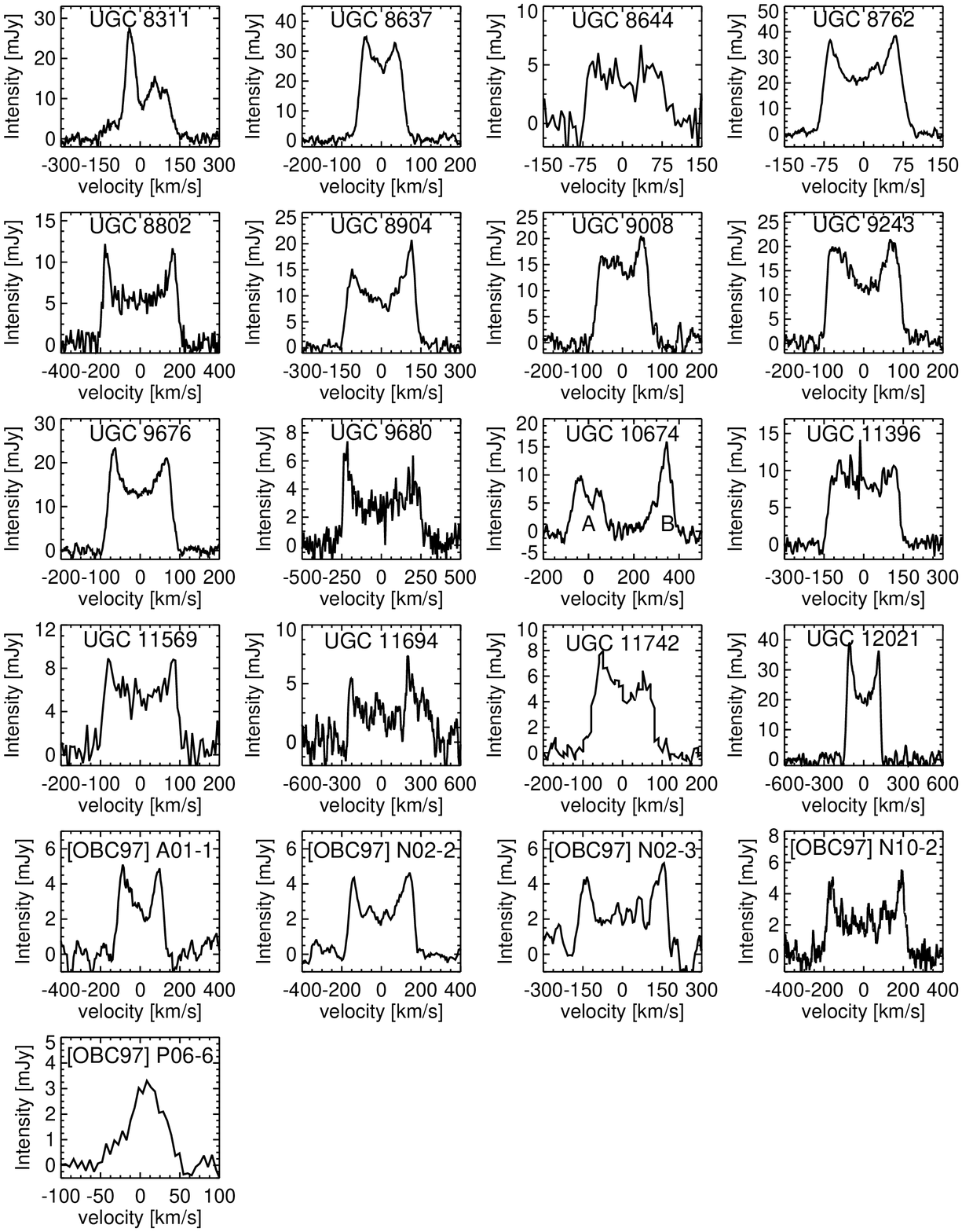}
\caption{{\it cont.} \HI\ spectra of all detected galaxies, excluding cases of possible confusion.
Channel resolution is given in Table~\ref{tab:obs}.}
\end{figure*}

\begin{figure*}
\centering
\includegraphics[width=6.0in]{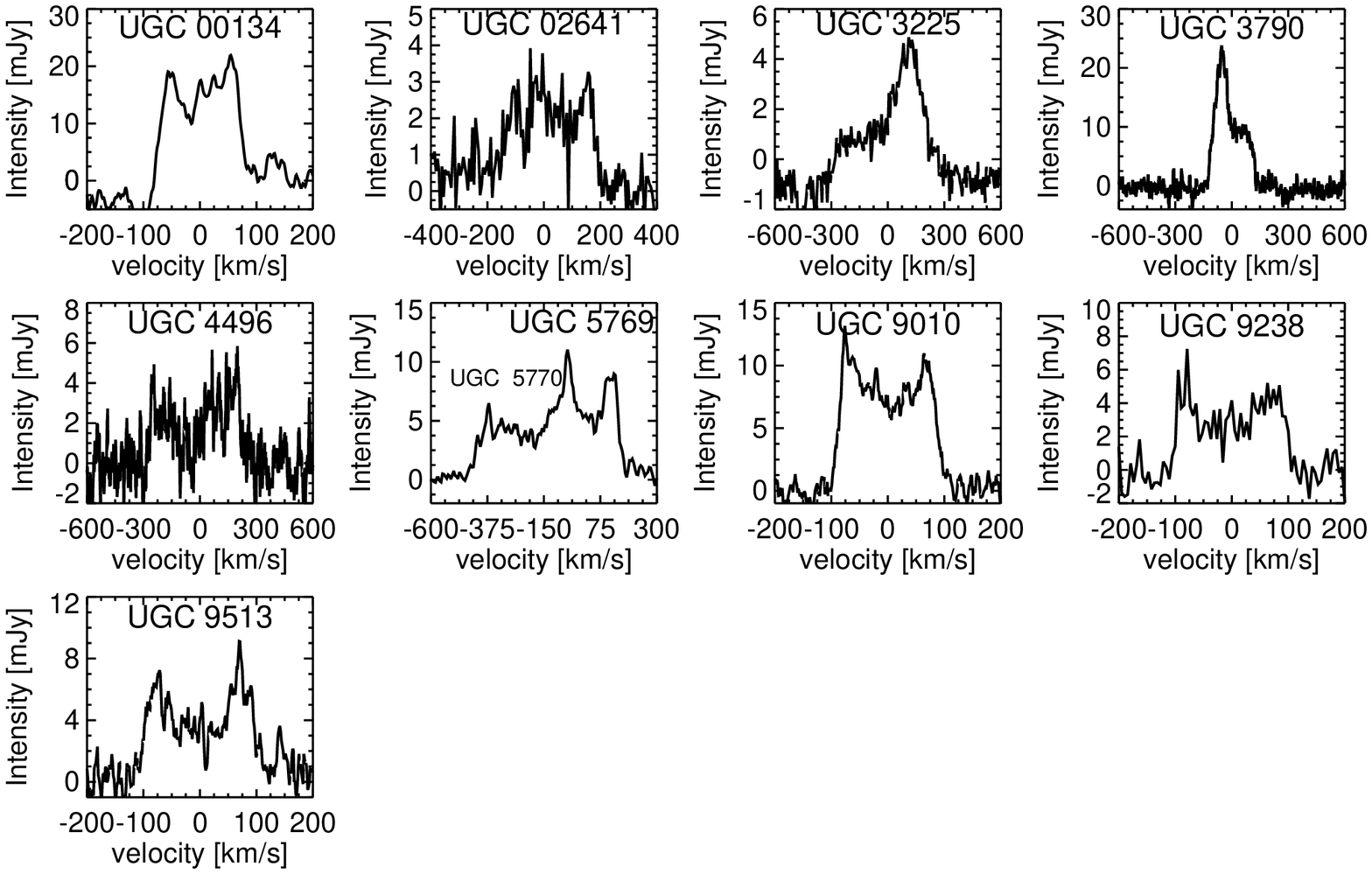}
\vskip -4.0in
\caption{\HI\ spectra of all detected galaxies which may suffer from confusion.
Channel resolution is given in Table~\ref{tab:obs}.}
\label{fig:hi_conf}
\end{figure*}

\subsection{Nan\c{c}ay Observations}

\begin{table}
\centering
\small 
\caption{Known Properties of Observed Galaxies \label{tab:props}} 
\begin{tabular}{lccclc}            
\hline\\[-10pt]
\hline \\
{\bf Galaxy}& {$\rm \mathbf m_{pg}$\dag} & {$\rm \mathbf \langle \mu_B \rangle$\ddag} &{$\rm \mathbf D_{25}$\dag}   
&{\bf Type\dag} & {\bf \it i\dag}\\ 
& &   & {$\left[\;\prime\;\right]$} &     
& {$\left[\;^\circ\;\right]$}            
\\
\hline\\             
UGC 00126  & 15.7 & 24.5 & 1.1 & S       & 26  \\
UGC 00134  & 17   & 25.6 & 1   & SIV-V   & 0   \\
UGC 00189  & 17   & 26.0 & 1.2 & ... & 43 \\
UGC 00266  & 16   & 24.8 & 1.1 & S?      & 50 \\
UGC 00293  & 17   & 25.8 & 1.1 & ... & 71\\
UGC 00424  & 18   & 26.6 & 1   & ... & 0 \\
UGC 00667  & 17   & 26.0 & 1.2 & ... & 50 \\
UGC 00795  & 16.5 & 25.1 & 1   & Sb        0 \\
UGC 01122  & 17   & 25.8 & 1.1 & Sc      & 35 \\
UGC 01362  & 18   & 25.6 & 1   & Dwrf Sp & 0\\
UGC 02299  & 17   & 25.6 & 1   & SA(s)dm & 27 \\
UGC 02580  & 18   & 26.6 & 1   & Pec     & 0 \\
UGC 02588  & 17   & 25.6 & 1   & Irr     & 0 \\
UGC 02641  & 17   & 26.0 & 1.2 & Sdm     & 65 \\
UGC 02671  & 17   & 26.0 & 1.2 & SAB(s)d & 0 \\
UGC 02741  & 16.5 & 25.7 & 1.3 & S?      & 0 \\
UGC 02796  & 16.5 & 26.0 & 1.5 & Sbc     & 73 \\
UGC 02797  & 16   & 25.0 & 1.2 & ... & 41 \\
UGC 02856  & 17   & 25.6 & 1   & Sb      & 71 \\
UGC 03119  & 16.5 & 25.5 & 1.2 & Sbc     & 69 \\
UGC 03225  & 16.5 & 25.3 & 1.1 & S       & 63 \\
UGC 03308  & 16.5 & 25.1 & 1   & Sc      & 0 \\
UGC 03585  & 14.7 & 23.3 & 1.0 & Sc      & 35 \\
UGC 03710  & 15.6 & 24.8 & 1.3 & ... & ... \\
UGC 03790  &...&...&0.8& SB?     & 67 \\ 
UGC 04109  & 14.5 & 23.3 & 1.1 & SB(r)b  & 24 \\
UGC 04131  & 15.7 & 25.1 & 1.4 & SBc     & 50 \\
UGC 04144  & 15.6 & 24.4 & 1.1 & Sc      & 83 \\
UGC 04288  & 17   & 25.6 & 1   & S0      & 37 \\
UGC 04496  & 15.7 & 25.8 & 2.0 & Sc      & 84 \\
UGC 04831  & 15.0 & 23.6 & 1.0 & SABd    & 0 \\
UGC 05009  & 14.8 & 23.6 & 1.1 & Scd     & 35 \\
UGC 05211  & 15.5 & 24.1 & 1.0 & ... & 0 \\
UGC 05361  & 16   & 24.6 & 1   & ... & 0 \\
UGC 05440  & 16.5 & 25.7 & 1.3 & Sd      & 65 \\
UGC 05583  & 16   & 24.6 & 1   & Sb      & 78 \\
UGC 05592  & 15.7 & 24.5 & 1.1 & Sc      & 43 \\
UGC 05679  & 15.4 & 25.2 & 1.7 & S       & 69 \\
UGC 05710  & 16   & 25.9 & 1.8 & S       & 74 \\
UGC 05743  & 16   & 24.8 & 1.1 & Sc      & 71 \\
UGC 05769  & 15.4 & 24.2 & 1.1 & ... & 25 \\
UGC 05770  & 15.4 & 24.0 & 1.0 & ... & 0 \\
UGC 05785  & 15.4 & 23.3 & 0.7 & S?      & 60 \\
UGC 05790  & 15.6 & 24.6 & 1.2 & S?      & 24 \\
UGC 05801  & 15.7 & 24.5 & 1.1 & ... & 0 \\
UGC 05828-1& 15.7 & 24.5 & 1.1 & Sc      & 0 \\ 
UGC 05828-2& 14.7 & 23.1 & 0.9 & S?      & 80 \\ 
UGC 05930  & 16.8 & 20.4 & 0.1 & S?      & 66 \\
UGC 06031  & 15.7 & 24.5 & 1.1 & S       & 0 \\
UGC 06124  & 16.5 & 26.0 & 1.5 & S       & 82 \\
UGC 06243  & 16   & 24.6 & 1   & Sc      & 46 \\
UGC 06300  & 15.7 & 24.7 & 1.2 & E       & 54 \\
UGC 06401  & 14.8 & 23.2 & 0.9 & Scd?    & 0 \\
UGC 06425  & 15.4 & 24.6 & 1.3 & Sb      & 23 \\
UGC 06524  & 11.6 & 24.8 & 8.1 & SB      & 63 \\
UGC 06525-1& 16.9 & 22.9 & 0.3 & S?      & 60 \\ 
UGC 06525-2& 16.0 & 22.6 & 0.4 & S?      & 73 \\    
\end{tabular}            
\end{table}
\addtocounter{table}{-1}
\begin{table}
\centering
\small
\caption{Known Properties of Observed Galaxies {\it cont.}}
\begin{tabular}{lccclc}            
\hline\\[-10pt]
\hline \\
{\bf Galaxy}& {$\rm \mathbf m_{pg}$\dag} & {$\rm \mathbf \langle \mu_B \rangle$\ddag} &{$\rm \mathbf D_{25}$\dag}   
&{\bf Type\dag} & {\bf \it i\dag}\\ 
& &   & {$\left[\;\prime\;\right]$} &     
& {$\left[\;^\circ\;\right]$}\\
\hline\\
UGC 06557  & 16   & 24.6 & 1   & Pec     & 26 \\
UGC 06559  & 15.7 & 25.4 & 1.7 & Sc      & 86 \\
UGC 06748  & 16   & 24.6 & 1   & S       & 37 \\
UGC 06842  & 16.5 & 25.5 & 1.2 & S       & 41 \\
UGC 06897  & 16   & 25.0 & 1.2 & SBc     & 65 \\
UGC 06913  & 15.3 & 25.0 & 1.6 & Sb      & 60 \\
UGC 06947  & 16.5 & 25.1 & 1   & Sc      & 84 \\
UGC 07084  & 14.9 & 24.6 & 1.6 & S       & 51 \\
UGC 07369  & 14.7 & 23.3 & 1.0 & E       & 0 \\
UGC 07388  & 16   & 24.6 & 1   & SB      & 37 \\
UGC 07425  & 16.5 & 25.5 & 1.2 & Sd      & 34 \\
UGC 07437  & 16   & 24.8 & 1.1 & SBc     & 60 \\
UGC 07438  & 16   & 25.5 & 1.5 & Sdm     & 75 \\
UGC 07457  & 16   & 25.0 & 1.2 & S       & 65 \\
UGC 07598  & 15.6 & 25.1 & 1.5 & SBc     & 0 \\
UGC 07630  & 16.5 & 26.0 & 1.5 & Dwrf Ir & 48 \\
UGC 07770  & 16.5 & 25.3 & 1.1 & Sc      & 84 \\
UGC 07928  & 14.4 & 23.0 & 1.0 & Sab     & 48 \\
UGC 07929  & 17   & 25.6 & 1   & Dwarf   & 46 \\
UGC 07934  & 16   & 25.0 & 1.2 & S       & 60 \\
UGC 08081  & 17   & 25.6 & 1   & Dwarf   & 72 \\
UGC 08171  & 15.3 & 23.7 & 0.9 & Sd      & 26 \\
UGC 08311  & 15.2 & 23.3 & 0.8 & S?      & 37 \\
UGC 08637  & 14.4 & 23.2 & 1.1 & SA(s)d  & 0 \\
UGC 08644  & 16   & 24.8 & 1.1 & Sdm     & 25 \\
UGC 08762  & 17   & 26.8 & 1.7 & SB(s)d  & 50 \\
UGC 08799  & 16.5 & 25.7 & 1.3 & Im?     & 48 \\
UGC 08802  & 17   & 26.2 & 1.3 & Sc      & 63 \\
UGC 08904  & 16   & 25.9 & 1.8 & ... & 44 \\
UGC 09008  & 14.6 & 23.2 & 1.0 & SAd     & 0 \\
UGC 09010  & 15.1 & 23.5 & 0.9 & Sd      & 37 \\
UGC 09238  & 16   & 25.2 & 1.3 & Sc      & 67 \\
UGC 09243  & 15.6 & 25.1 & 1.5 & Sc      & 58 \\
UGC 09513  & 15.4 & 24.2 & 1.1 & ... & 35 \\
UGC 09676  & 15.7 & 24.5 & 1.1 & SBc     & 43 \\
UGC 09680  & 16.5 & 25.7 & 1.3 & Scd     & 81 \\
UGC 09767  & 14.2 & 22.8 & 1.0 & BCG     & 26 \\
UGC 09770  & 14.9 & 23.7 & 1.1 & SABb    & 43 \\
UGC 10217  & 16.0 & 24.6 & 1.0 & SBb     & 36 \\
UGC 10365  & 16.0 & 24.6 & 1.0 & S       & 73 \\
UGC 10377  & 17   & 26.4 & 1.4 & Pec?    & 69 \\
UGC 10673  & 16   & 25.7 & 1.6 & ... & 64 \\
UGC 10674-1& 15.5 & 25.3 & 1.7 & S       & 79 \\ 
UGC 10674-2& 15.5 & 25.3 & 1.7 & S       & 79 \\ 
UGC 11396  & 17   & 25.6 & 1   & ... & 60 \\
UGC 11569  & 16   & 24.8 & 1.1 & Sc      & 51 \\
UGC 11625  & 16.5 & 25.5 & 1.2 & S       & 24 \\
UGC 11654  & 16.5 & 25.1 & 1   & Pec     & 60 \\
UGC 11694  & 15.2 & 25.1 & 1.8 & ... & 52 \\
UGC 11742  & 16.5 & 25.5 & 1.2 & S       & 0 \\
UGC 11840  & 17   & 25.4 & 0.9 & S?      & 27 \\
UGC 12021  & 15.0 & 24.0 & 1.2 & Sb      & 65 \\
UGC 12189  & 16.0 & 24.4 & 0.9 & SB?     & 48 \\
UGC 12359  & 16.5 & 26.0 & 1.5 & S?      & 21 \\
UGC 12424  & 15.2 & 24.0 & 1.1 & S       & 35 \\
\obc\ P06-6& 18.4 & 23.9 & 0.3 & S       & 33 \\  
\obc\ N02-2& 17.0 & 22.4 & 1.0 & S       & 69 \\  
\end{tabular}            
\end{table}
\addtocounter{table}{-1}
\begin{table}
\centering
\small
\caption{Known Properties of Observed Galaxies {\it cont.}}
\begin{tabular}{lccclc}            
\hline\\[-10pt]
\hline \\
{\bf Galaxy}& {$\rm \mathbf m_{pg}$\dag} & {$\rm \mathbf \langle \mu_B \rangle$\ddag} &{$\rm \mathbf D_{25}$\dag}   
&{\bf Type\dag} & {\bf \it i\dag}\\ 
& &   & {$\left[\;\prime\;\right]$} &     
& {$\left[\;^\circ\;\right]$}\\
\hline\\
\obc\ N02-3& 18.2 & 22.2 & 0.4 & S       & 27 \\  
\obc\ N10-2& 16.6 & 22.3 & 0.5 & S       & 64 \\  
\obc\ A01-1& 17.9 & 23.1 & 0.6 & S       & 47 \\  
\hline\\
\multicolumn{6}{l}{\dag From the UGC (Nilson 1973).}\\
\multicolumn{6}{l}{\ddag Surface brightness ($\rm mag\;arcsec^{-2}$) is $\mu=m_{pg}+5\log{(D)}$}\\
\multicolumn{6}{l}{\hskip 0.1in $+8.89-0.26$; where $m_{pg}$ is the photographic magnitude}\\
\multicolumn{6}{l}{\hskip 0.1in from the UGC, D is the diameter in arcmin, 8.89 is the}\\
\multicolumn{6}{l}{\hskip 0.1in conversion from arcmin to arcsec, and 0.26 is an average}\\
\multicolumn{6}{l}{\hskip 0.1in conversion from m$_{pg}$ to m$_B$.}        \\
\hline\\[-10pt]
\hline\\     
\end{tabular}            
\end{table}             

The Nan\c{c}ay decimetric radio telescope, a meridian transit-type instrument
of the Kraus/Ohio State design, consists of a fixed spherical mirror (300~m long
and 35~m high), a tilt-able flat mirror (200$\times$40~m), and a focal carriage 
moving along a curved rail track. Sources on the celestial equator can be tracked 
for approximately 60 minutes. The telescope's collecting area is about 7000~m$^{2}$ 
(equivalent to a 94-m diameter parabolic dish). 
Due to the E-W elongated shape of the mirrors, some of the instrument's
characteristics depend on the declination at which one observes. At 21-cm 
wavelength the telescope's half-power beam width (HPBW) is \am{3}{5} in right 
ascension, independent of  declination, while in the north-south direction it is 23$'$ 
for declinations up to $\sim$20$^{\circ}$, rising to 25$'$ at $\delta$=
40$^{\circ}$ (see also Matthews \& van Driel \cite{matthews00}). Although the instrument's effective collecting area 
and, consequently, its gain, follow the same geometric effect, decreasing 
correspondingly with declination, this effect is negligible for the declination range
of the objects in our sample.
All observations for our project were made after a major renovation of the
instrument's focal system (e.g., van Driel et al. \cite{vandriel97})
which resulted in a typical system temperature of 35~K.

The observations were made in the period of January -- December 2002. 
using a total of 59 hours of telescope time.
We obtained our observations in total power (position-switching) mode
using consecutive pairs of 40 seconds on and 40 seconds 
off-source integrations. Off-source integrations were taken at a position
about 20$'$~E of the target position.

Only galaxies with published redshifts were observed at Nan\c{c}ay.
For all observations the autocorrelator was divided into 1 pair
of cross-polarized receiver banks, each with 4096 channels and a 25~MHz 
bandpass, resulting in a channel spacing of 1.3~\kms. The center 
frequencies of the 2 banks were tuned to the redshifted \HI\ line
frequency of the target source.
These spectra were boxcar smoothed to a channel separation of 17.1~\kms\
during the data reduction in order to increase signal-to-noise.  

Flux calibration, i.e., the conversion of observed system temperatures to
flux densities in mJy, is determined for the Nan\c{c}ay telescope through
regular measurements of a cold load calibrator and periodic monitoring of strong
continuum sources by the Nan\c{c}ay staff. Standard calibration procedures include
correction for the above mentioned declination-dependent gain variations of the 
telescope (e.g., Fouqu\'e et al. \cite{fouque90}). Additionally, a number of standard
calibrator galaxies were observed throughout  our observing runs, showing the gain 
used to be within 10\%.

In order to reduce the effect of radio frequency interference (RFI) in our
observations, we used an off-line RFI mitigation program, which is part of 
the standard NAPS software package, see Monnier Ragaigne et al. (2003) 
for further details. It should be noted, however, that not all unwanted
emissions could be elimination using this method and that residual signals
frequently occur around 4700, 8300, 9500, 11300, 12700 and 13600 \kms\
(Monnier-Ragaigne et al. \cite{monnier03}).

\begin{figure*}
\centerline{
\includegraphics[width=3.0in]{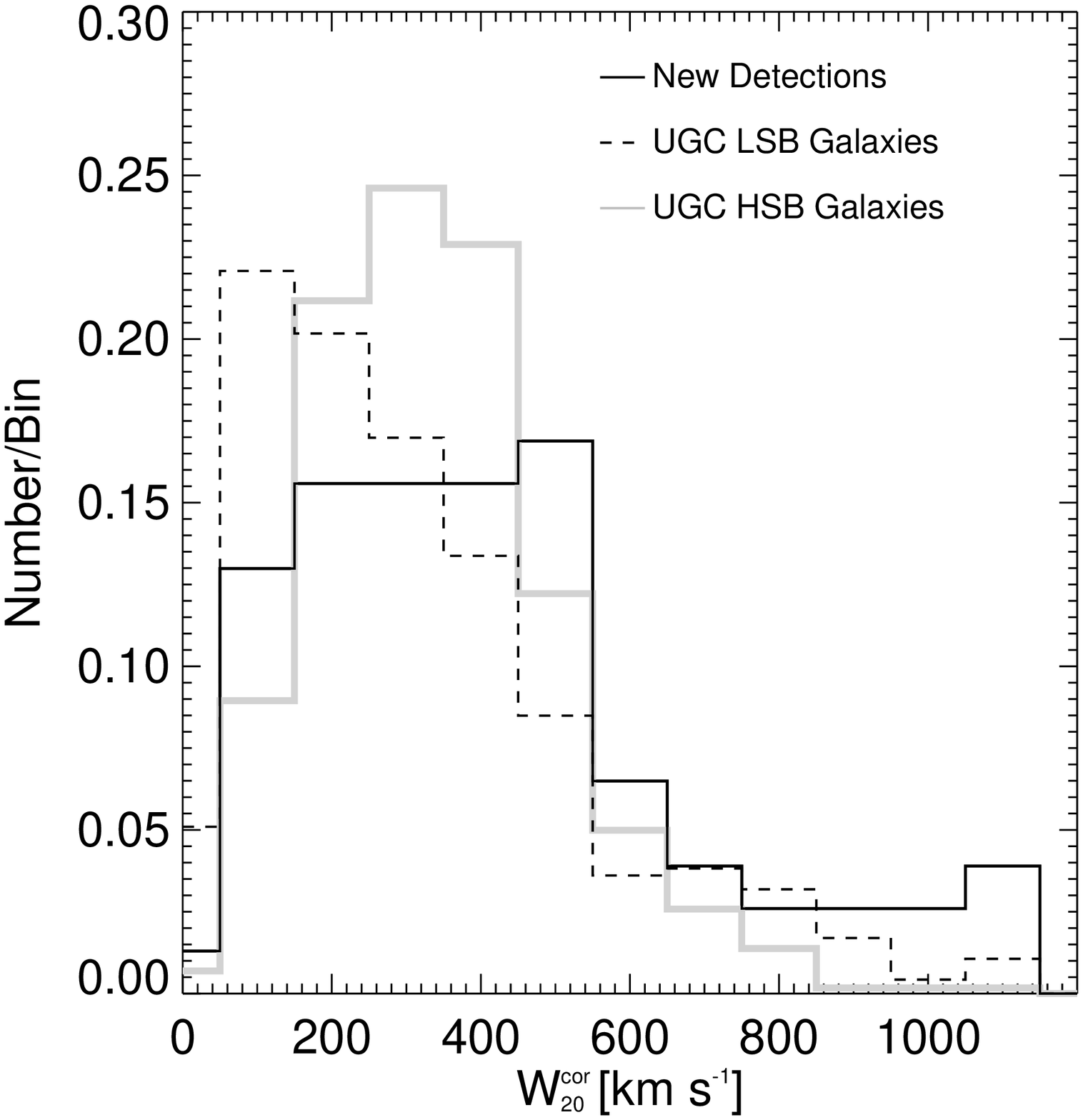}
\includegraphics[width=3.0in]{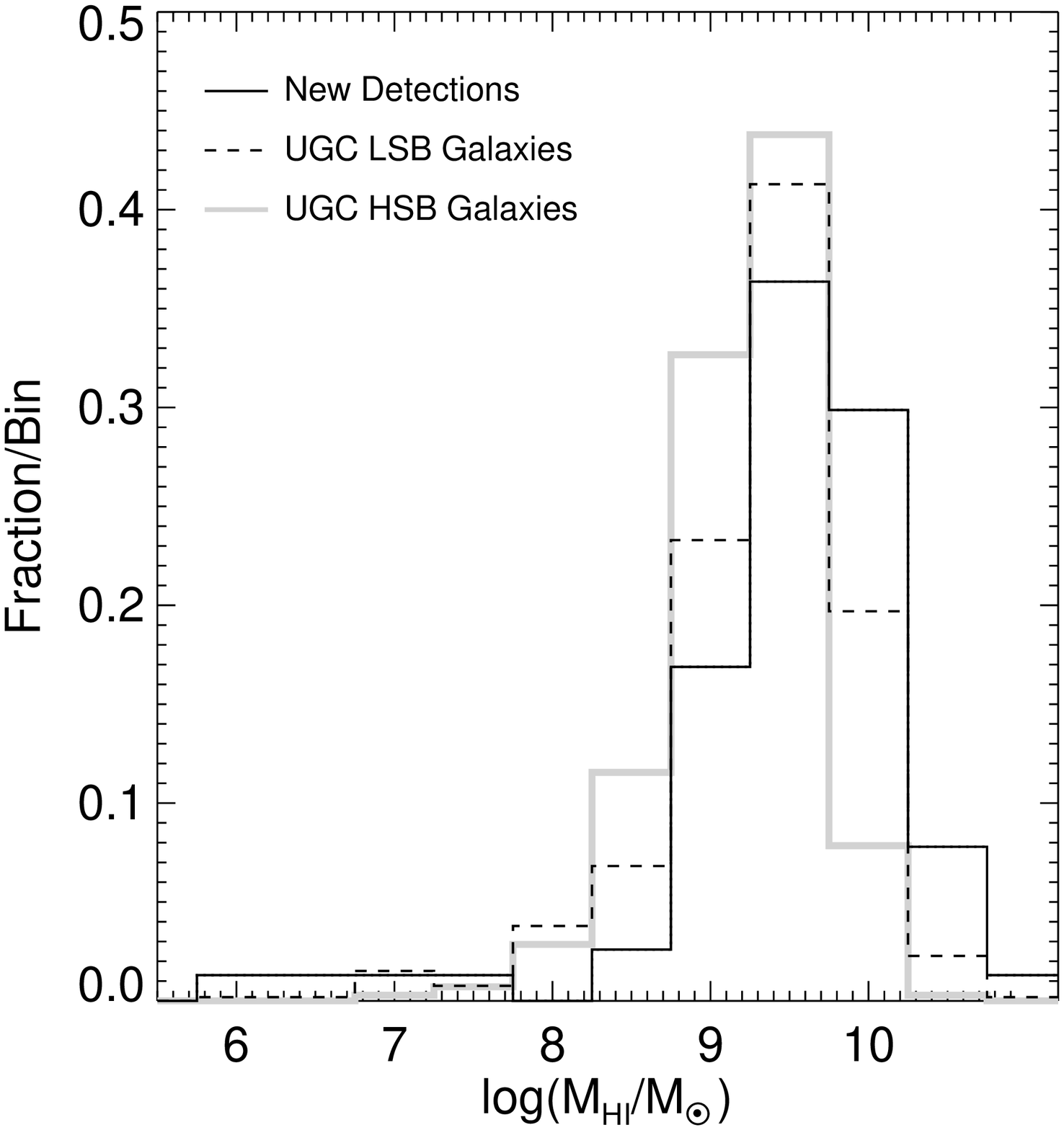}}
\caption{The fractional distribution of 
velocity widths (at 20\% of the peak, and corrected for inclination, left) and
\HI\ masses (right) for all the galaxies in this survey. 
In both plots, the new detections are shown by the solid black line
while the results from all LSB UGC galaxies in this study are shown as the dashed line.
Additionally, the results from a similar \HI\ survey done by Giovanelli \& Haynes (\cite{giov93}) of HSB CGCG galaxies is shown by the thick gray line.
In the plot on the left, the inclination correction applied is simply W$_{20}^{corr}$ = W$_{20}$/sin({\it i}).  To avoid over-correction, any inclination
less than 30$^\circ$ has been set to 30$^\circ$ for the purpose of this correction.
Note that the extremely high values of W$_{20}^{corr}$ may be due to an underestimate 
of the galaxy's inclination (see Figure~\ref{fig:W20i}).}
\label{fig:hist_mass}
\end{figure*}

\begin{figure}
\centering
\includegraphics[width=3.75in]{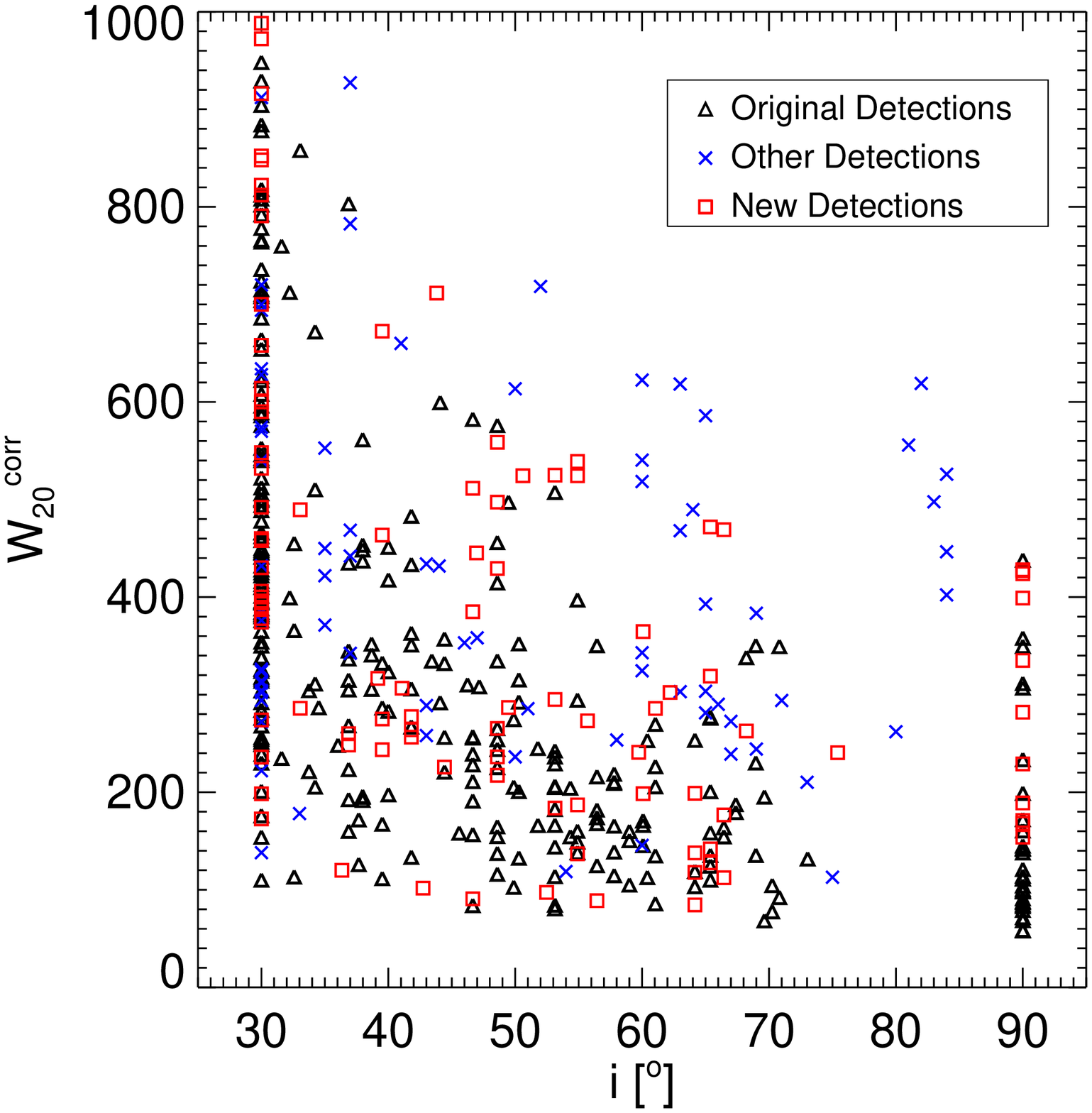}
\caption{The inclination of each LSB galaxy in our survey
versus its inclination-corrected velocity width.
From this plot it can be seen that the extremely high values of W$_{20}^{corr}$
are found when the inclinations are the lowest.  This indicates that 
the inclination in those cases may be underestimated, artificially
increasing the values for  W$_{20}^{corr}$.
The inclination plotted is that given by the UGC catalog, except in 
the cases where {\it i}$<$30\degree.   To avoid over-correction, any inclination
less than 30$^\circ$ has been set to 30$^\circ$ for the purpose of the 
velocity width correction.}
\label{fig:W20i}
\end{figure}

\section{Observational results}

Results of the Arecibo and Nan\c{c}ay observations are given in Tables~\ref{tab:obs}
and \ref{tab:hi_props} and the spectra of all detected galaxies are given in Figures~\ref{fig:hi_spectra}
and Figures~\ref{fig:hi_conf}.  
Notes on the individual galaxies can be found in Appendix A.

Table~\ref{tab:obs} lists all galaxies observed,
as well as the search range, resolution, telescope used, RFI encountered, and the r.m.s. of
the RFI-free regions of the spectra.  Table~\ref{tab:hi_props} lists all observed galaxies
which were detected in \HI.  Table~\ref{tab:hi_props} is laid out as follows:
\begin{itemize}
\item {\bf Column 1:} The galaxy name;
\item {\bf Column 2:} The heliocentric velocity as measured by our observations;
\item {\bf Column 3 \& 4:} The {\it uncorrected} velocity widths at 20\% and 50\% of 
the peak or average height of the two peaks (when applicable);
\item {\bf Column 5:} The total measured \HI\ line flux of the galaxy;
\item {\bf Column 6:} The total \HI\ mass of the galaxy, assuming H$_0$=75 \kms\ Mpc$^{-1}$ 
and obtained using the standard formula
$\rm M_{HI}/M_{\odot} = 2.356\times10^5 \left({{v[km\; s^{-1}]}\over{H_0 [km\;  s^{-1}\; Mpc^{-1}]}}\right)^2 \int{S_\nu[Jy] d\nu} $
;
\item {\bf Column 7:} All previously measured velocities of the galaxy, as
found in the literature;
\item {\bf Column  8:} Any notes on the observation regarding potential confusion
with other sources.  Notes are given in Appendix A.
\end{itemize}

Figures~\ref{fig:MHID} and \ref{fig:W20D} show the distribution of the \HI\ mass
and corrected \HI\ line width for the objects in the survey against distance.
For comparison, the plots also included all other massive LSB galaxies found in the literature, as well
as three other Arecibo \HI\ surveys -- the 'blind' \HI\ surveys of Zwaan et. al (\cite{zwaan97}) and Rosenberg
\& Schneider (\cite{rosenberg00}), and a similar survey to look for high surface brightness CGCG galaxies done
by Giovanelli \& Haynes(\cite{giov93}).

\begin{figure}
\centering
\includegraphics[width=3.75in]{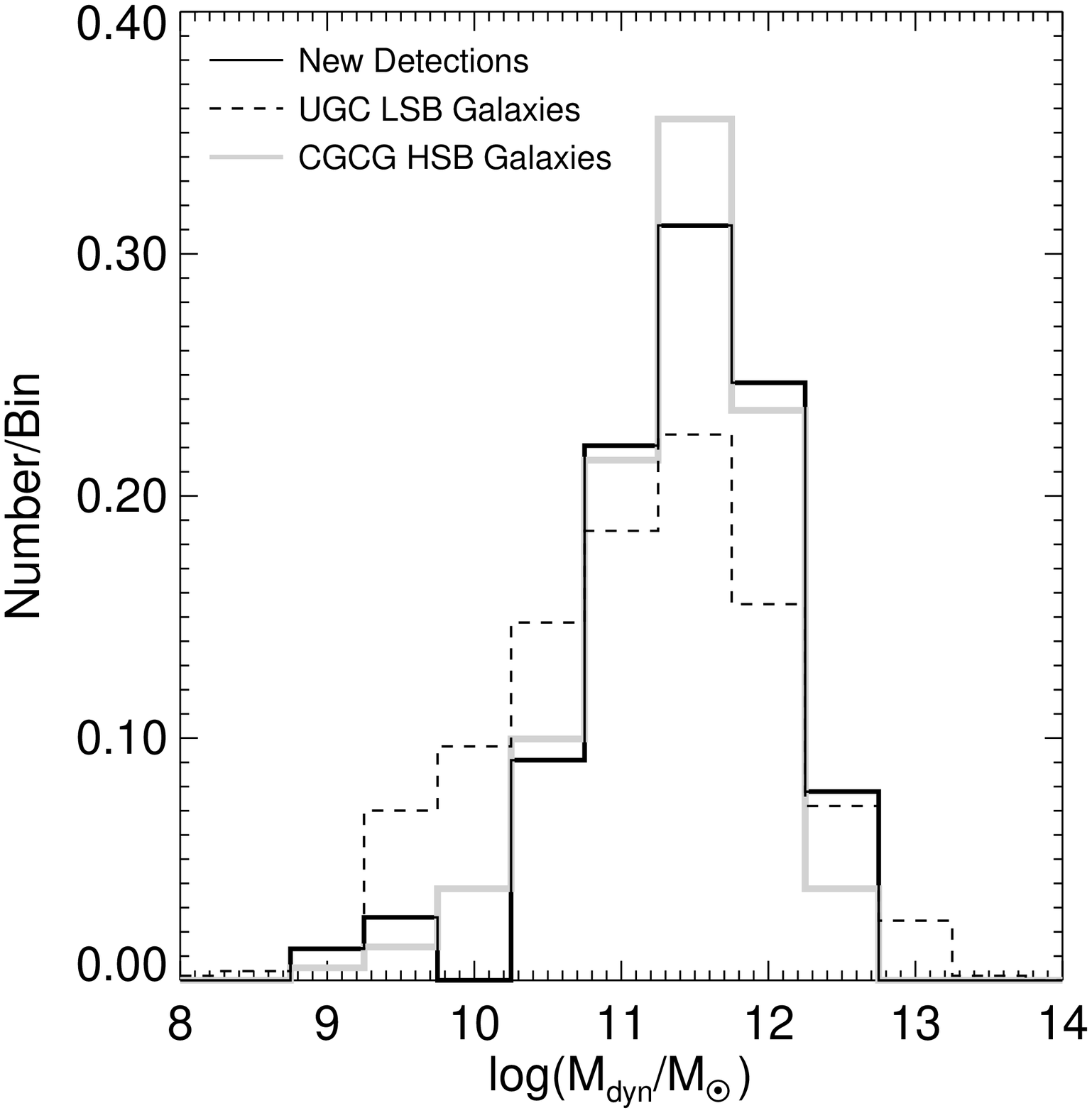}
\caption{The fractional distribution of dynamical mass for all the galaxies in this survey.
In both plots, the new detections are shown by the solid black line
while the results from all LSB UGC galaxies in this study are shown as the dashed line.
Additionally, the results from a similar survey done by
Giovanelli \& Haynes (\cite{giov93}) looking for \HI\ in the HSB galaxies of the CGCG is
shown by the thick gray line.}
\label{fig:hist_mass2}
\end{figure}

\section{\HI\ and Stars in LSB Galaxies}

\begin{figure*}
\centerline{
\includegraphics[width=3.0in]{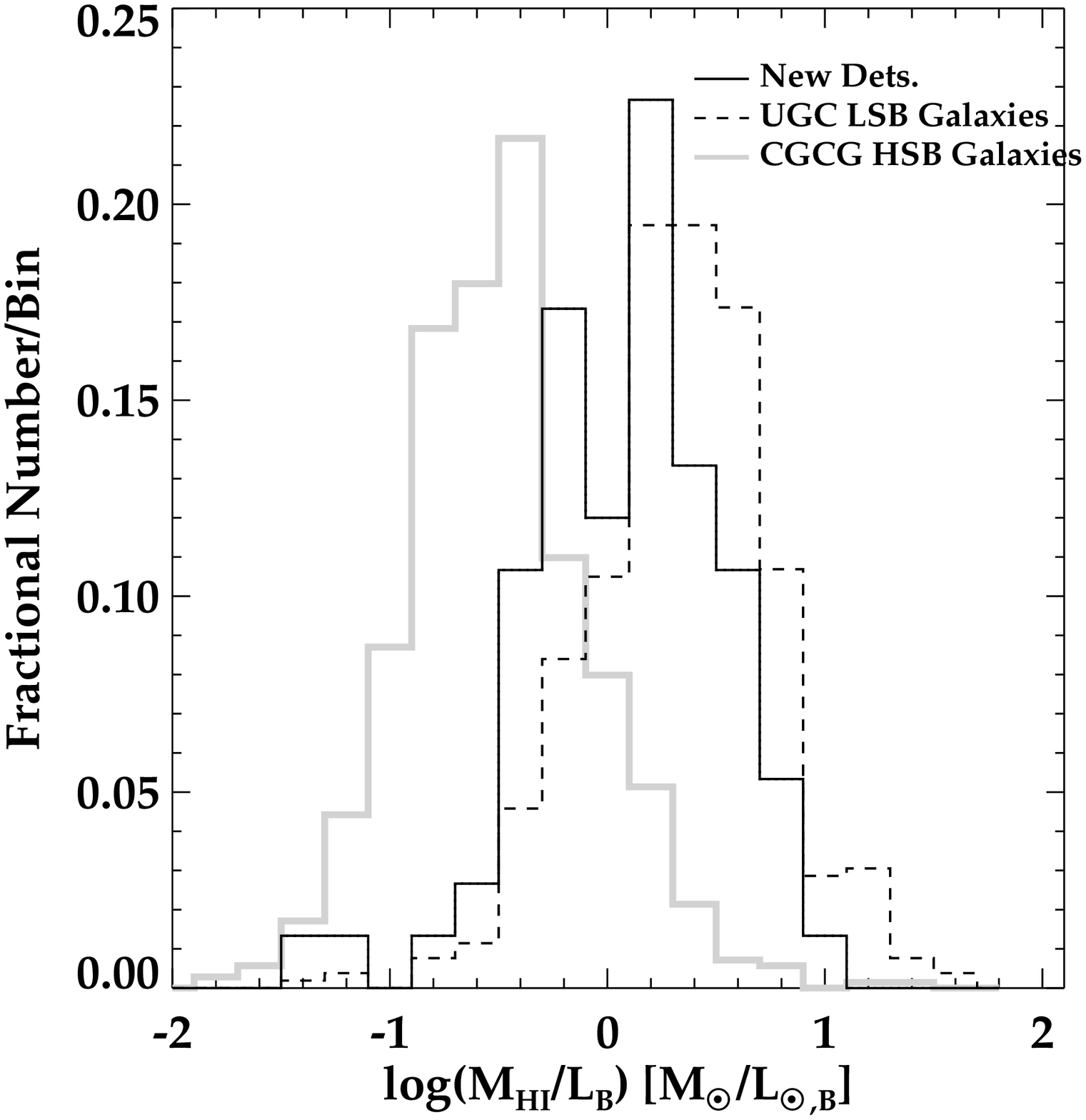}
\includegraphics[width=3.0in]{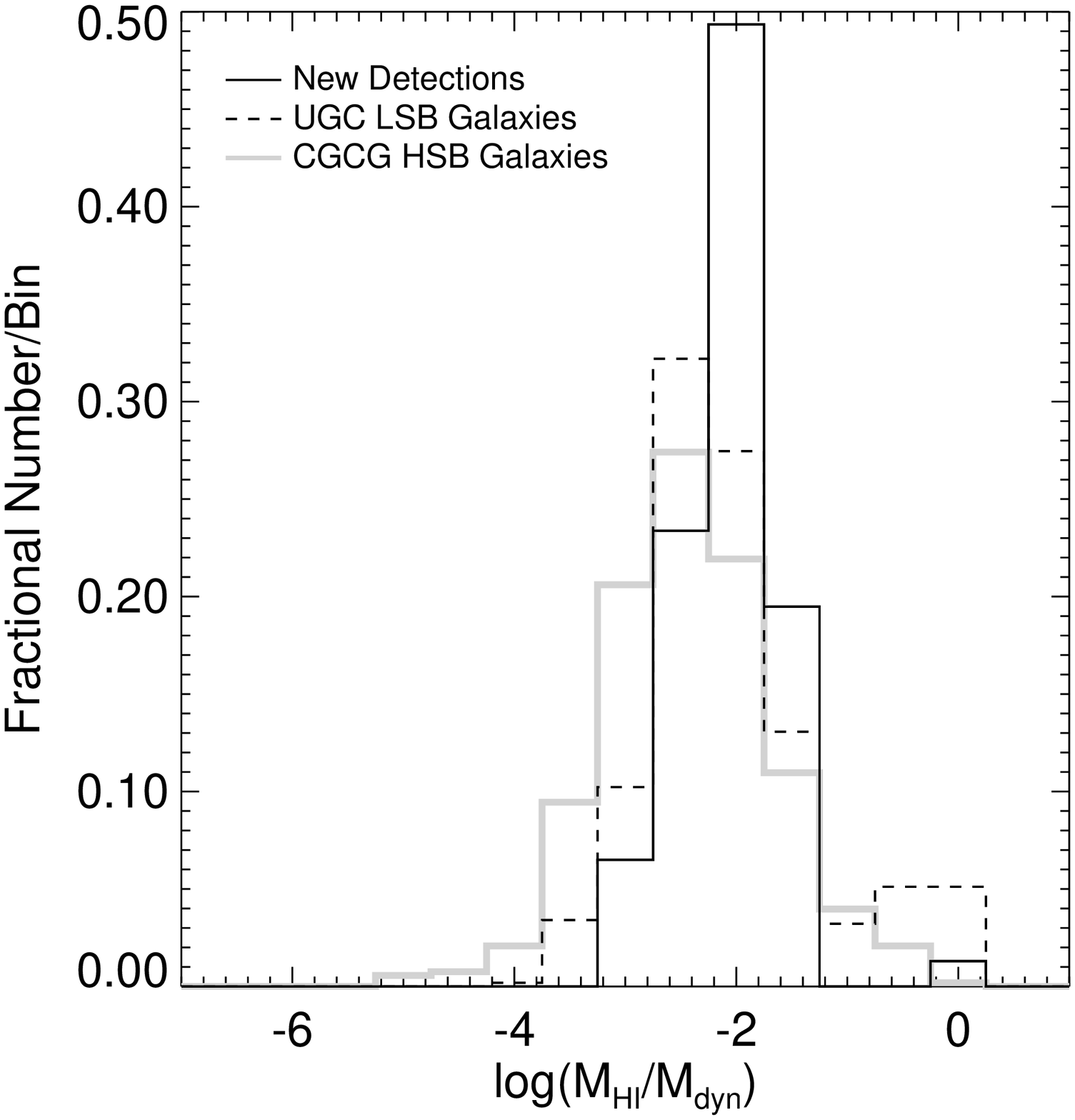}}
\caption{The fractional cumulative distribution of \HI\ mass-to-luminosity ratio (left) and
\HI\ mass-to-dynamical mass (right) for the UGC LSB galaxies.
In both plots, the new detections are shown by the solid black line
while the results from all LSB UGC galaxies in this study are shown as the dashed line.
Additionally, the results from a similar survey done by
Giovanelli \& Haynes (\cite{giov93}) looking for \HI\ in the HSB galaxies of the CGCG is
shown by the thick gray line.}
\label{fig:hist_mass3}
\end{figure*}

\begin{figure}
\centering
\includegraphics[width=3.75in]{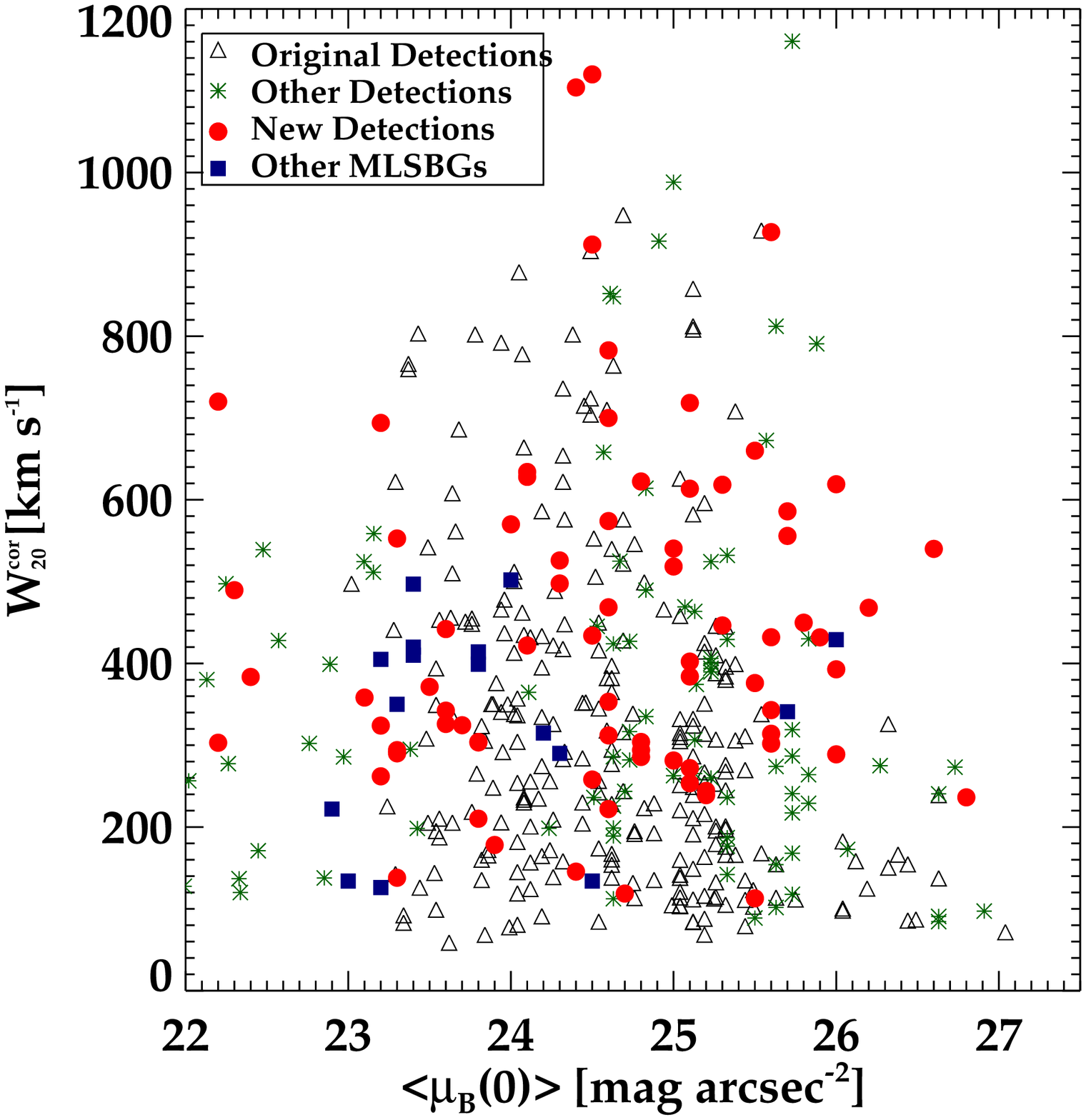}
\caption{Surface brightness versus (inclination corrected) velocity widths both for all the 
galaxies in our LSB UGC sample and for the 16 massive LSB galaxies with cataloged
\HI\ properties (Matthews et al. \cite{matthews01};
Sprayberry et al. \cite{sprayberry95}).  
The inclination correction applied is simply
W$_{20}^{corr}$ = W$_{20}$/sin({\it i}).  To avoid over-correction, any inclination
less than 30$^\circ$ has been set to 30$^\circ$ for the purpose of this correction.
Note that the extremely high values of W$_{20}^{corr}$ may be due to an underestimate
of the galaxy's inclination (see Figure~\ref{fig:W20i}).}
\label{fig:mu_W20}
\end{figure}

\begin{figure*}
\centerline{
\includegraphics[width=3.0in]{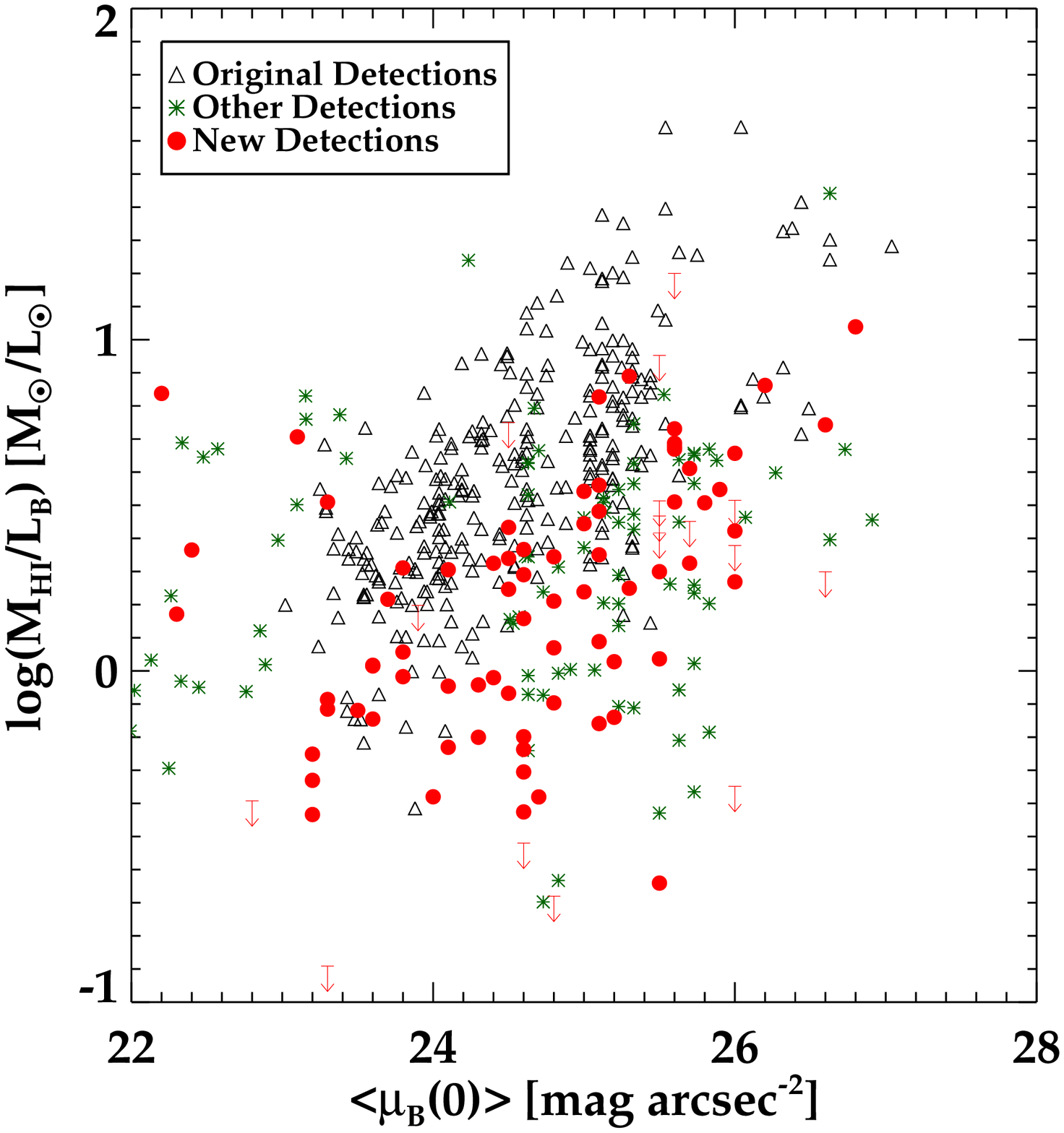}
\includegraphics[width=3.0in]{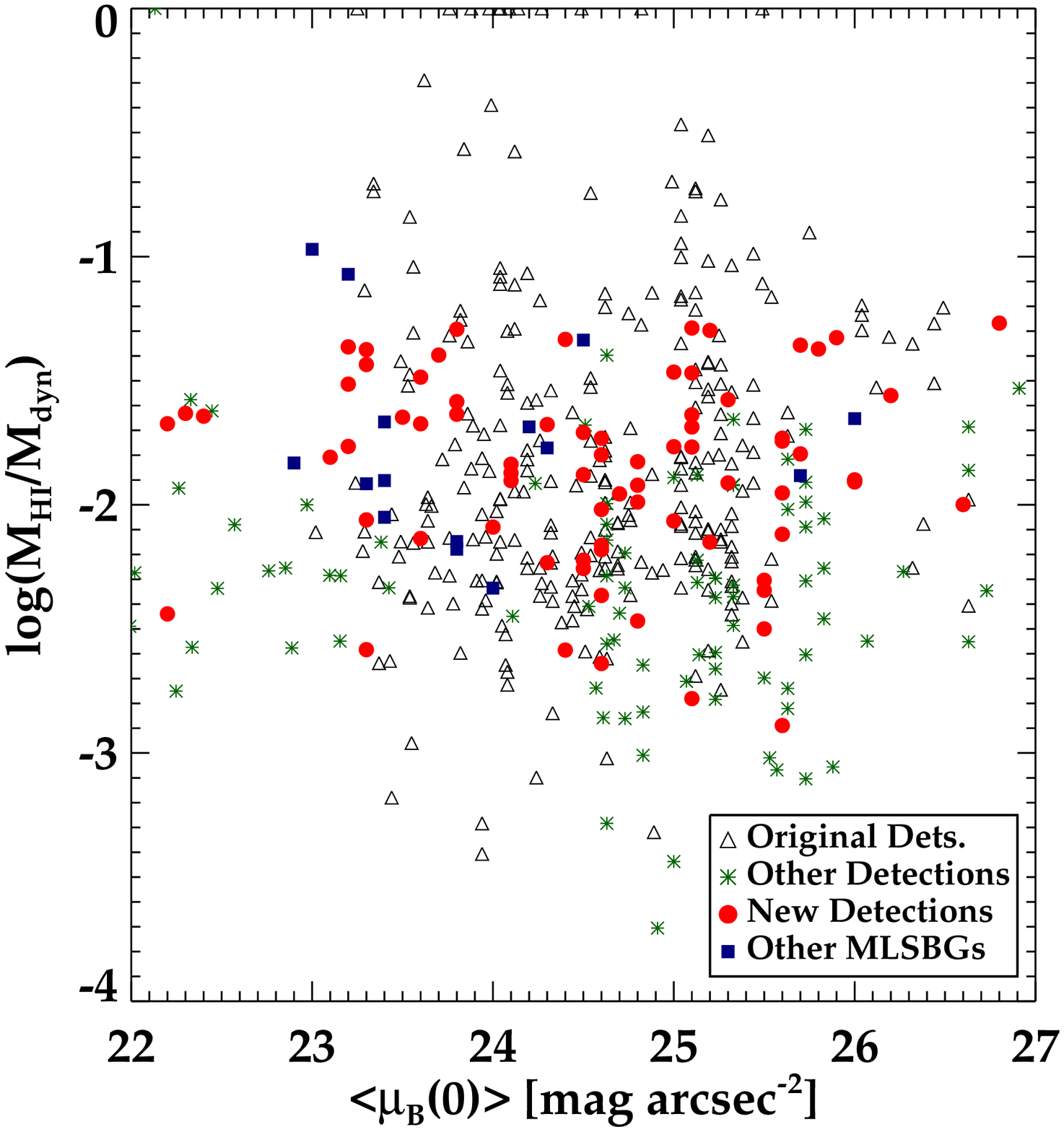}}
\caption{Surface brightness versus  M$_{HI}$/L$_B$ (left)
and M$_{HI}$/M$_{dyn}$ (right) for all the galaxies in our LSB UGC sample, as well
as the 16 massive LSB galaxies with published \HI\ properties (Matthews, van Driel,
\& Monnier-Ragaigne \cite{matthews01}).
Those objects which were observed in our survey which have published velocities but
which were not detected are given upper limits to their flux equal to 
3$\sigma \times \langle W_{20} \rangle$, where $\langle W_{20} \rangle$~=~300~\kms.
These objects are shown as arrows in the left plot.}
\label{fig:mu_M}
\end{figure*}

\begin{figure*}
\centerline{
\includegraphics[width=3.0in]{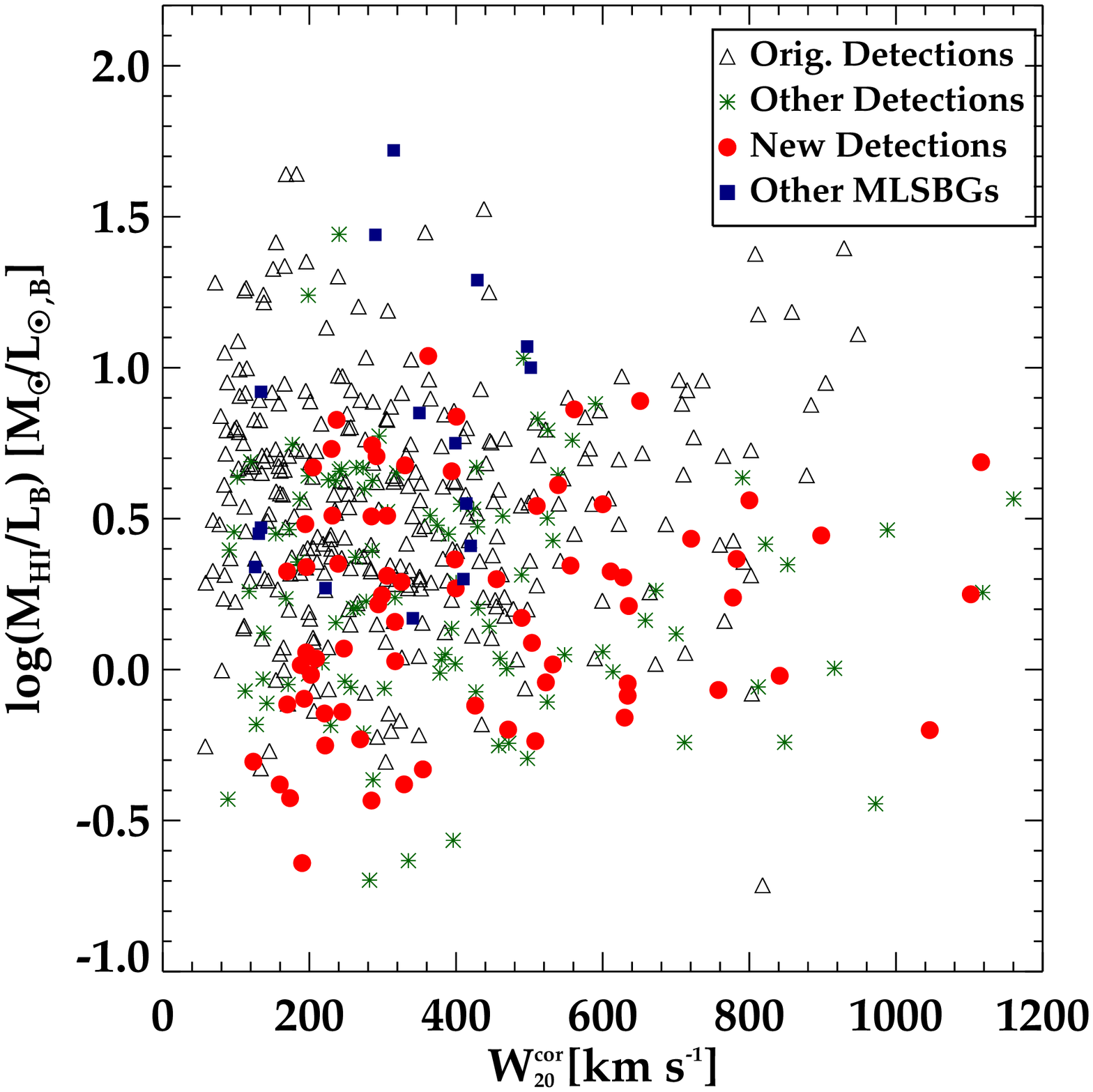}
\includegraphics[width=3.0in]{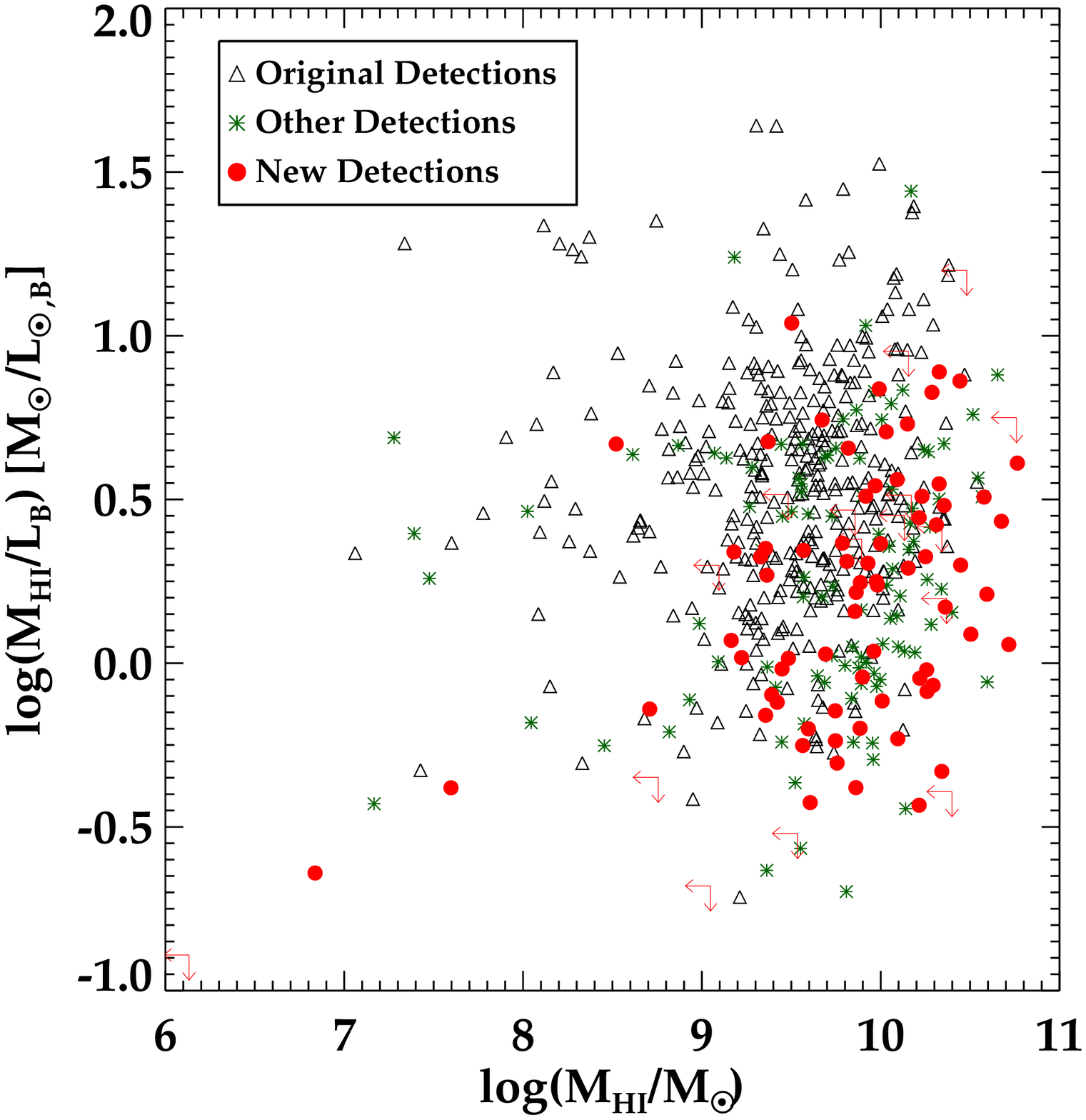}}
\caption{Velocity width (corrected for inclination) (left) and \HI\ gas mass (right) 
versus M$_{HI}$/L$_B$ for all the galaxies in our LSB UGC sample. 
Those objects which were observed in our survey which have published velocities but
which were not detected are given upper limits to their flux equal to 
3$\sigma \times \langle W_{20} \rangle$, where $\langle W_{20} \rangle$~=~300~\kms.
These objects are shown as arrows in the plots.
In the plot on the left, the inclination correction applied is simply
W$_{20}^{corr}$ = W$_{20}$/sin({\it i}).  To avoid over-correction, any inclination
less than 30$^\circ$ has been set to 30$^\circ$ for the purpose of this correction.
Note that the extremely high values of W$_{20}^{corr}$ may be due to an underestimate
of the galaxy's inclination (see Figure~\ref{fig:W20i}).
In addition, the extremely high values for M$_{HI}$/L$_B$ are likely due
to underestimates of L$_B$ in the UGC catalog, a result of the galaxies'
LSB nature.}
\label{fig:ML}
\end{figure*}

Figures~\ref{fig:hist_mass} -- \ref{fig:ML} show the distribution of the \HI\
and optical properties of the galaxies in the Bothun et al. (\cite{bothun85}) catalog for which we  
have new measurements.  From these figures it is clear that the dynamic range
of physical properties of LSB disks is large.
That is, the UGC LSB 
galaxy catalog described herein has objects with \HI\ masses ranging from 
$\rm 6 \le log(M_{HI}/M_\odot) \le 11$, observed velocity widths of $\rm 55 \le W_{20} \le 638$~\kms\
($\rm 58 \le W_{20}^{cor} \le 1276$), and \HI\ mass-to-luminosity ratio within the range
$\rm 0.04 \le M_{HI}/L_B \le 44$ \Msol/\Lsol. 
Thus, we have a catalog of LSB disk galaxies which range from dwarf through
massive galaxies, and include objects which are both gas-rich and gas-poor.

Examining Figures~\ref{fig:hist_mass} -- \ref{fig:hist_mass3} it is clear that
while our UGC LSB galaxy sample has similar gas and dynamical masses to that found with
HSB galaxies, the M$_{HI}$/L$_{B}$ ratios in the LSB galaxies of this survey are 
much higher than in typical HSB galaxies.
More specifically, the median \HI\ mass of the galaxies in this survey is 
$4\times10^9 $ \Msol, close to the median of 3$\times10^9$ \Msol\ found
in a similar survey of HSB galaxies from the {\it Catalog of Galaxies
and Clusters of Galaxies} (CGCG) (Zwicky et al. \cite{zwicky61}) done by Giovanelli \& Haynes (\cite{giov93}).
The dynamical masses are also similar,
with $\rm \langle M_{dyn}\rangle_{median}^{LSB} = 3\times10^{11}\;and\; 
\langle M_{dyn}\rangle_{median}^{HSB} = 5\times10^{11}$. 
Yet the range of mass-to-luminosity ratios for the two samples of galaxies is quite
different.  The median $\rm M_{HI}/L_B$ value for the LSB galaxies is 2.8 \Msol/\Lsol\
while $\rm \langle M_{HI}/L_B\rangle_{median}^{HSB}$ = 0.4 \Msol/\Lsol.  The gas-to-dynamical
mass ratio shows a similar, albeit smaller, trend with $\rm \langle M_{HI}/M_{dyn}\rangle_{median}^{LSB} = 0.01\;
and \;\langle M_{HI}/M_{dyn}\rangle_{median}^{HSB} = 0.006$.

The higher \HI\ mass-to-luminosity ratios (M$_{HI}$/L$_{B}$)
found for the LSB galaxies may be a strong indicator
that the LSB systems have evolved differently from their  HSB counterparts.
This idea is by no means a new one -- that LSB galaxies in general have high \HI\ mass-to-luminosity 
ratios is well established.  In \cite{romanishin82}, Romanishin et al. first reported that galaxies of lower surface
brightness have $\rm M_{HI}/L_B$ values approximately 2.4 times higher than for similar
HSB samples.  This trend has continued for all \HI\ studies undertaken of these diffuse
systems (e.g. de Blok et al. \cite{deblok96}; O'Neil et al. \cite{oneil00}; 
Schombert et al. \cite{schombert01}; Burkholder et al. \cite{burkholder01}) 
indicating that LSB
galaxies are less evolved, less efficient in their star formation, or have otherwise
undergone a different evolutionary scenario then their HSB counterparts 
(see, e.g., the modeling by Boissier et al. \cite{boissier03}).

However, this brings us to a logical paradox that was first explored by
de Blok and McGaugh (\cite{deblok98}).  Our calculations of dynamical mass (e.g.
$v^2R$) use the tabulated UGC diameter as the basis for calculating R.
In general this diameter corresponds to an
isophotal limit of 25 \mss\ (Cornell, et.al \cite{cornell87}). For a typical HSB galaxies with \Bmo = 21.5,
this isophotal radius corresponds to $\sim$3 optical disk scale lengths. 
However, for a typical LSB galaxy with \Bmo = 23.5, this
isophotal radius is only 1-2 scale lengths.  This means that for given
values of circular velocity, redshift, and apparent diameter, an LSB
will have a dynamical mass which is systematically underestimated compared
to that of an HSB galaxy.  We estimate that this error could be as high as 
a factor of two or more.
Therefore, if a constant number of scale lengths were used
in the determination of the radius,
the fractional \HI\ content of LSB disks would actually be
lower than that of HSB disks.  This is another way of saying that LSB disks must
have a lower baryonic mass fraction than HSB disks.  Similar conclusions,
based on mass modeling of LSB galaxies with rotation curves, have been
reached by McGaugh \& de Blok (\cite{mcgaugh98}) and Pickering et al. (\cite{pickering99}).   This lower
baryonic mass fraction could result in different star
formation histories between LSB and HSB disks.

Indeed, Figure~\ref{fig:mu_M}(a) shows a curious correlation between
$\langle\mu_B\rangle$ and $\rm M_{HI}/L_B$,
similar to that found by, e.g. Burkholder et al. (\cite{burkholder01})  -- an increasing \HI\ mass-to-luminosity
ratio with decreasing surface brightness.   Although there is a large
scatter, the trend seen in this figure is significant.  In particular,
the mean value of $\rm M_{HI}/L_B$ for $\langle\mu_B\rangle$ = 23.0, 24.0 and 25.0
is 0.24, 0.37, and 0.55, respectively.  To first order, these trends are physically reasonable.  For
instance, a purely gaseous disk sitting in some dark matter potential
will have very low optical surface brightness and a very high
gas to star ratio.   However, that simplistic scenario would also predict
a correlation between fractional gas content and surface brightness.  That is,
as star formation begins to consume the gas and thus elevate the surface
brightness of the disk, the fractional H I content of the potential would
diminish. One should then observe the lowest values of $\langle\mu_B\rangle$
corresponding to the highest values of $\rm M_{HI}/M_{dyn}$. However,
this is clearly not seen as these two parameters are essentially uncorrelated (Figure~\ref{fig:mu_M}b).

Thus, while the average gas-to-light ratio of galaxies appears to 
increase considerably with decreasing surface brightness, the average
gas-to-{\it total} mass ratio does not show any increase.
Indeed, if the dynamical masses of the lower surface brightness galaxies
are underestimated, as discussed above, using accurate measurements of
M$_{dyn}$ could result in the inverse -- a {\it decrease} of $\rm M_{HI}/M_{dyn}$
with $\langle\mu_B\rangle$.
If the only factor causing the $\rm M_{HI}/L_B\;-\;\mu_B$ correlation
were a slower or less efficient evolution of LSB systems, the correlation should be
equally evident between $\langle\mu_B\rangle$ and both $\rm M_{HI}/M_{dyn}$ and $\rm M_{HI}/L_B$.
As this is not the case,
additional possibilities for reducing the luminosity of a galaxy with surface brightness while
keeping the gas-to-total mass ratio the same must be considered. This is
where the physical meaning of surface brightness comes into play, as surface
brightness is the convolution of the mean (blue) luminosity of the stellar
population and the average separation between the stars.  One explanation
of these trends is that the process of star formation in LSB disks is sufficiently
different than that seen in HSB disks so as to reduce the natural coupling
between increased blue luminosity and increased blue surface brightness.
This can be accomplished (as has been suggested earlier by O'Neil et al. \cite{oneil00} and
Bothun et al. \cite{bothun97}) if star formation in LSB disks occurs in a lower density
gas environment which produces a larger than average separation between 
newly formed stars.

\begin{figure*}
\centerline{
\includegraphics[width=3.0in]{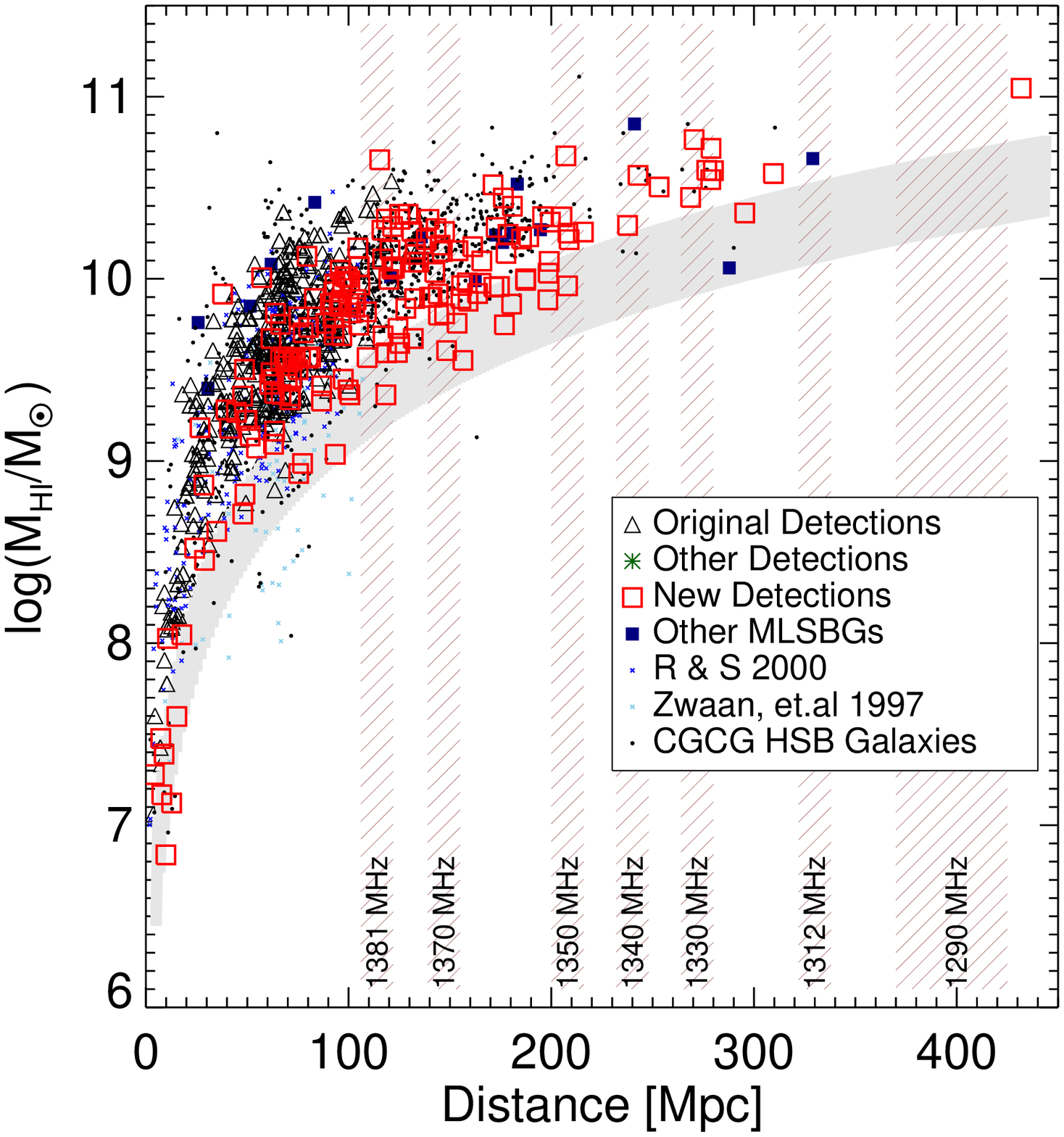}
\includegraphics[width=3.0in]{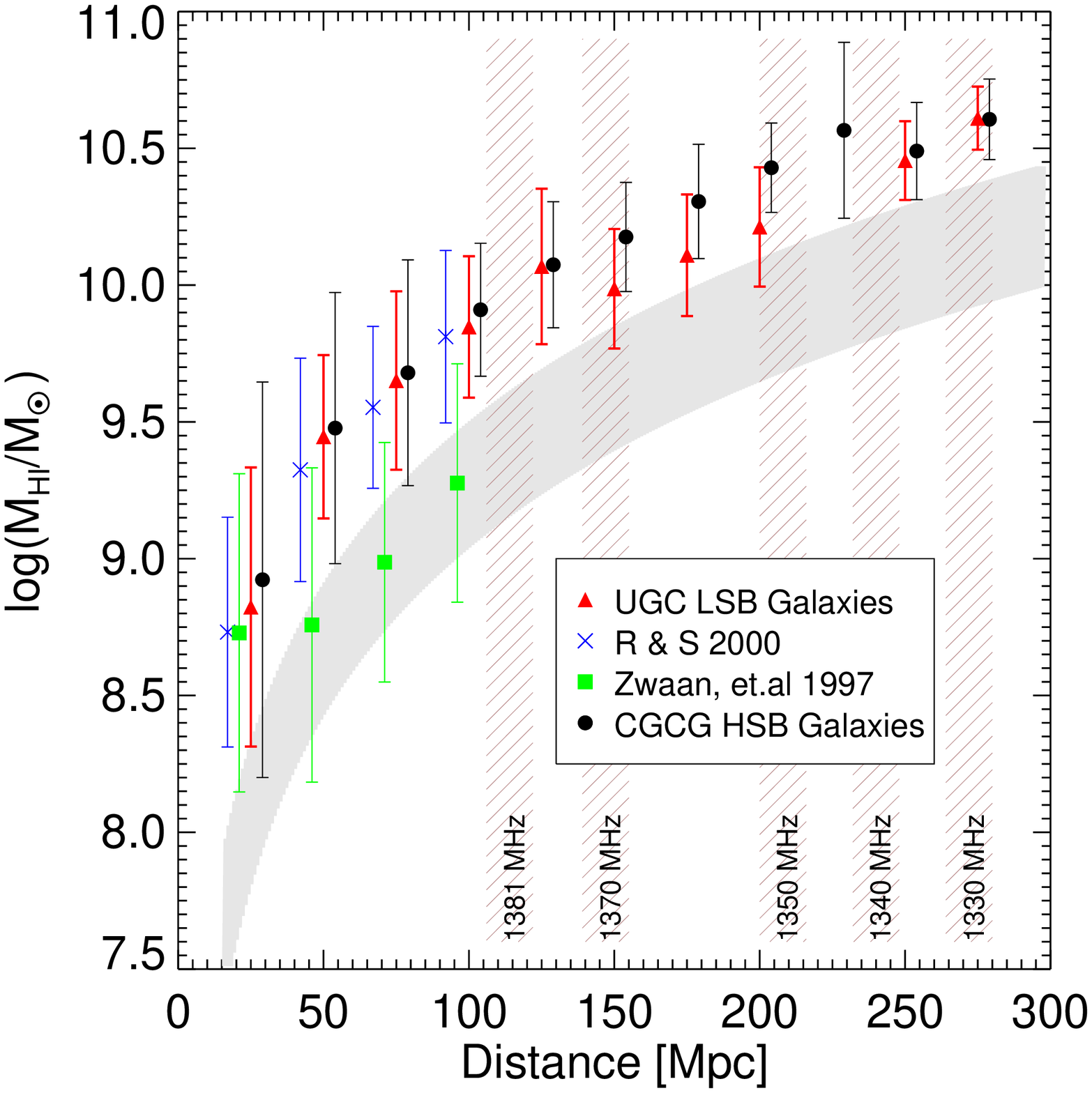}}
\caption{Distance (${velocity}\over{H_0}$) plotted against total \HI\ mass for
all the galaxies in the total UGC LSB galaxy discussed herein (Section 2).
For comparison, objects found in the two Arecibo blind
\HI\ surveys (Rosenberg \& Schneider \cite{rosenberg00}; Zwaan et al. \cite{zwaan97}) and
the Giovanelli \& Haynes (\cite{giov93}) CGCG survey are also plotted.  The gray line on the
plots indicates the approximate survey limit for our observations, where the line thickness
represents the varying sensitivities and velocity widths found herein.  Finally, the
diagonal lines show the primary regions affected by RFI.  On the left the individual data points are
shown, and the 16 other massive LSB galaxies with published \HI\ masses  (from Matthews et al. \cite{matthews01}
and Sprayberry et al. \cite{sprayberry95}) are also plotted.  On the right, the mean values for the various surveys
are plotted, with 25 Mpc bins.  Note the data points in each bin are offset in the in the x-direction
for clarity -- the points should lie over the red symbols.}
\label{fig:MHID}
\end{figure*}

\begin{figure*}
\centerline{
\includegraphics[width=3.0in]{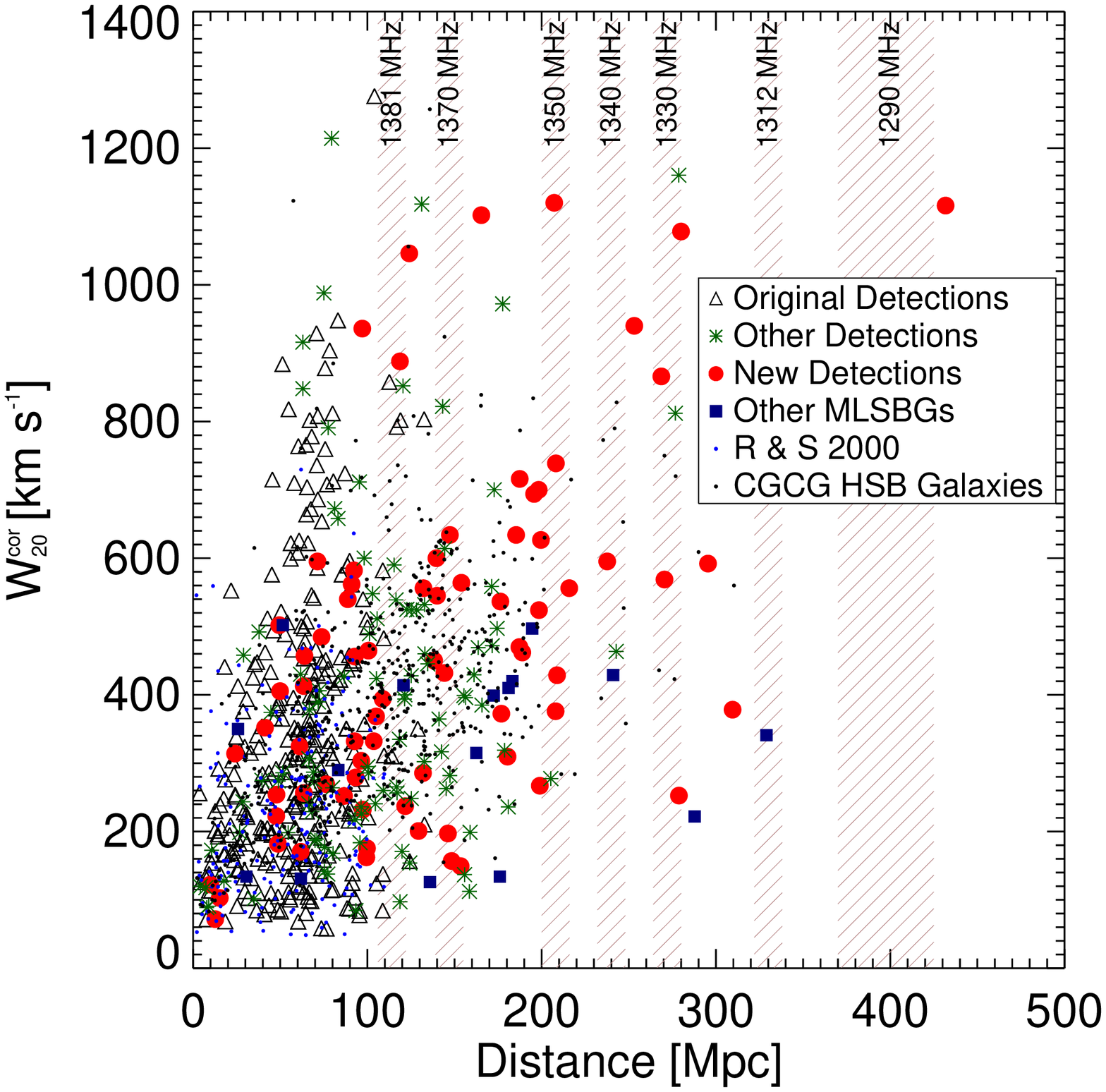}
\includegraphics[width=3.0in]{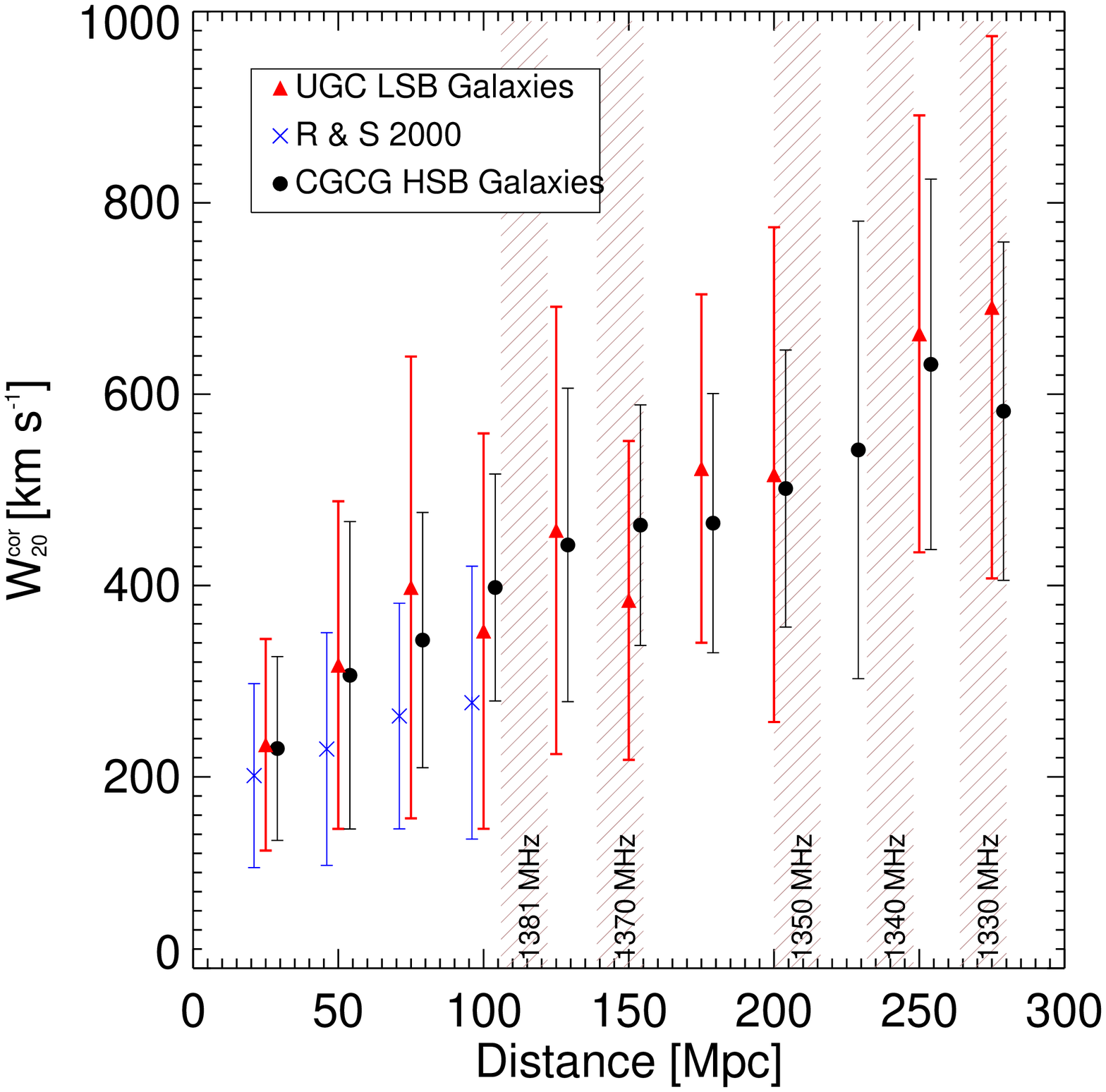}}
\caption{Distance (${velocity}\over{H_0}$) plotted against velocity width
(corrected for inclination) for
all the UGC LSB galaxies discussed herein (Section 2).
For comparison, objects found in the Rosenberg \& Schneider (\cite{rosenberg00}) Arecibo blind
\HI\ survey and the Giovanelli \& Haynes (\cite{giov93}) CGCG survey are also plotted.
(The Zwaan et al. \cite{zwaan97} survey did not publish velocity widths.)
Again, the diagonal lines show the primary regions affected by RFI.
On the left the individual data points are
shown, and the 16 other massive LSB galaxies with published \HI\ masses  (from Matthews et al. \cite{matthews01}
and Sprayberry et al. \cite{sprayberry95}) are also plotted.  On the right, the mean values for the various surveys
are plotted, with 25 Mpc bins.  Note the data points are offset in the in the x-direction
for clarity -- the points should lie over the red symbols.
In the plot on the left, the inclination correction applied is simply
W$_{20}^{corr}$ = W$_{20}$/sin({\it i}).  To avoid over-correction, any inclination
less than 30$^\circ$ has been set to 30$^\circ$ for the purpose of this correction.
Note that the extremely high values of W$_{20}^{corr}$ may be due to an underestimate
of the galaxy's inclination (see Figure~\ref{fig:W20i}).  }
\label{fig:W20D}
\end{figure*}
\section{Conclusions \& Discussion}

Our survey, undertaken with both the Arecibo and Nan\c{c}ay radio telescopes,
has resulted in the determination of \ion{H}{I} properties and redshifts for 81
UGC LSB galaxies, 38 of which can be classified as massive `Malin 1 cousins'.  
This has increased the number of known massive LSB galaxies by a factor of 3.  
Combining our results with all previous 21-cm observations of the Bothun et al. 
(\cite{bothun85}) UGC LSB galaxy list results in a final catalog of 526 LSB galaxies, 
with properties ranging across the known gamut of disk system properties, including 
numerous extremely gas-rich galaxies with M$_{HI}/L_B \ge$ 10 \Msol/\Lsol. Comparing 
the properties of these galaxies with those found in similar \HI\ surveys of HSB 
galaxies surveys shows the curious fact that while the mean M$_{HI}$/L$_B$ value 
increases with decreasing surface brightness, the M$_{HI}$/M$_{dyn}$ ratio does not. 
This raises the (currently unanswered) question of whether or not star formation 
is significantly affected by the low density environment found within LSB galaxies.

With this catalog in hand, we can now ask the question -- 
What do we gain by continuing to look for and study 
massive LSB systems?  The answer to this question is (at least) threefold.

First, as massive LSB galaxies are significant repositories of baryonic (and dark) matter 
(Pickering et al. 1997, 1999; Walsh et al. \cite{walsh97}; Pickering et al. \cite{pickering97}),
determining both the number density and distribution of these objects could provide considerable 
insight into the overall distribution of matter in the Universe.   While the survey described herein
cannot determine the number density of massive LSB galaxies, it does show that there are more of these
objects in the Universe than previously known, and determining the density and distribution 
of these massive objects will allow for a better understanding of the cosmological distribution of 
mass throughout the universe.

In this respect, it is vital here to point out the effect surveys such as this one have on 
determining the LSB galaxy contribution to baryon density. Because of their perceived morphology 
on photographic plates lacking the sensitivity to show their outer disks, and their previous lack 
of detection in \ion{H}{I}, all of the massive LSB galaxies detected in our catalog were previously 
believed to be dwarf systems with little to no \ion{H}{I} mass.
As a result, our detections are not merely adding a few massive LSB galaxies to 
the number counts, but moving galaxies from being listed as extremely low mass to 
extremely high mass systems, and thus shifting the distribution of masses in the luminosity function.

Second, and equally important, massive LSB galaxies provide a unique insight 
into galaxy evolution studies as a whole.  Because massive LSB galaxies reside
at an extreme end of the mass and density distribution of galaxies, any model 
that wishes to describe galaxy evolution must be able to account for them.  
And, as massive LSB galaxies appear to be undeveloped in at least some aspects, 
when compared to HSB galaxies (e.g., their higher than average M$_{HI}$/L$_B$ 
ratios and often sparse H-$\alpha$ emission -- Pickering et al. \cite{pickering97}), 
studies of massive LSB galaxies, and of LSB galaxies in general, offer a look into a 
different aspect of galaxy formation than those of HSB galaxies.

Third, it has been shown that if LSB galaxies are as common as HSB galaxies, 
then they may be one of the primary causes for Lyman-$\alpha$ absorbers
in the local Universe (Linder 1998, 2000). As a result, knowing the density of 
massive LSB galaxies could lead to a allow for considerably better understanding 
of the nature of Lyman-$\alpha$ absorption lines. This would allow us to construct 
a more complete picture of the large-scale gas distribution in the Universe, as well 
as providing insight into galaxy formation processes.


At this point it is clear that we do not yet have an answer regarding the number
density of massive LSB galaxies, as the blind surveys done to date 
(both optical, near-infrared and \HI) have not yet probed
to sufficient sensitivity to allow this question to be answered. Clearly there is a dearth of such
objects in the z$<$0.01 Universe, just as there is a shortage of all massive galaxies
in that region.  What is significant, though, is that as increasingly sensitive instruments
become available, the number of LSB and massive LSB galaxies known continues to increase.
(Notably, in a survey similar to ours, Schwortz et al. (\cite{schwortz04}) 
found another 9 massive LSB galaxies.)  However, as the nature of pointed surveys such as
the one described herein prevents them from providing an unbiased view into the number density
of various galaxy types, we can only use these surveys to get a glimpse into what may be
out there.  The true number density of massive LSB galaxies will likely not be known until
an extremely sensitive, large scale survey is undertaken, such as may be possible with
the Arecibo L-band Feed Array (ALFA) or, ultimately, the Square Kilometre Array (SKA).

\acknowledgements{ 
The Arecibo Observatory is part of the National Astronomy and Ionosphere Center, 
which is operated by Cornell University under a cooperative agreement with the National 
Science Foundation.  The \nan\ radio astronomy station, which is the Unit\'e Scientifique 
de \nan\ of the Observatoire de Paris, is associated with the French Centre National de 
Recherche Scientifique (CNRS) as USR B704, and acknowledges the financial support of the 
R\'egion Centre as well as of the European Union. This research has made use of the 
Lyon-Meudon Extragalactic Database (LEDA), recently incorporated in HyperLeda, and of 
the NASA/IPAC Extragalactic Database (NED) which is operated by the Jet Propulsion 
Laboratory, California Institute of Technology, under contract with the National 
Aeronautics and Space  Administration.
}

\begin{table*}
\centering
\caption{Observed Galaxies \label{tab:obs}}
\begin{tabular}{llllcccl}
\hline\\[-10pt]
\hline\\
{\bf Galaxy}& {\bf RA} & {\bf Dec} & {\bf Search Range}
&{$\sigma$} & {\bf resolution} &{\bf Tel.} &{\bf RFI \dag}\\
& {$\left[J2000\right]$} & {$\left[J2000\right]$}&
{$\left[km s^{-1}\right]$} &{$\left[mJy\right]$} &{$\left[km\;s^{-1}\right]$} \\
\hline\\
UGC 00126&  00:13:44.7& 14:36:25& 31,745 -- -970&  0.42& 26& A& o,p,q,s\\
UGC 00134&  00:14:04  & 12:53:46& 3,297 -- -1769&   3.2& 17& N& ...\\
UGC 00189&  00:19:58  & 15:05:31& 10,151 -- 5,085&  1.7& 17& N& ...\\
UGC 00266&  00:27:22.8& 10:50:54& 31,745 -- -970&  0.48& 26& A& o,q\\
UGC 00293&  00:29:19.8& 26:24:35& 31,745 -- -970&  0.77& 26& A& o,p,q,s\\
UGC 00424&  00:39:50.1& 20:31:04& 31,745 -- -970&  0.62& 26& A& h,o,p,q\\
UGC 00667&  01:04:58  & 05:39:10& 8,619 -- 3,553&   1.9& 17& N& ...\\ 
UGC 00795&  01:14:56.3& 12:22:51& 31,745 -- -970&  0.41& 26& A& o,p,q,s,t\\
UGC 01122&  01:34:19.6& 29:15:52& 31,745 -- -970&  0.42& 26& A& h,o,q\\
UGC 01362&  01:53:51  & 14:45:51& 10,451 -- 5,385& 0.46& 17& N& ...\\
UGC 02299&  02:49:08  & 11:07:11& 12,786 -- 7,720&  2.8& 17& N& ...\\
UGC 02580&  03:11:33.4& 06:42:16& 31,745 -- -970&  0.36& 26& A& h,o,p,q,s,t\\
UGC 02588&  03:12:26.7& 14:24:30& 31,745 -- -970&   1.2& 26& A& h,j,k,l,m,o,q,t\\
UGC 02641&  03:17:12.8& 03:35:58& 31,745 -- -970&  0.35& 26& A& h,o,p,q,r,s\\
UGC 02671&  03:19:24  & 08:07:24& 9,660 -- 4,594&  0.26& 17& N& ...\\
UGC 02741&  03:26:49  & 07:07:29& 13,535 -- 8,469&  2.6& 17& N& ...\\
UGC 02796&  03:36:53  & 13:24:24& 11,609 -- 6,543&  2.2& 17& N& ...\\
UGC 02797&  03:37:33.6& 23:17:35& 31,745 -- -970&  0.51& 26& A& h,o,p,q,r,s\\
UGC 02856&  03:46:30  & 15:24:39& 11,266 -- 6,200&  9.2& 17& N& ...\\
UGC 03119&  04:39:08  & 11:31:49& 10,384 -- 5,318&  2.7& 17& N& ...\\
UGC 03225&  04:59:24.5& 12:46:00& 31,745 -- -970&  0.36& 26& A& o,p,q,s\\
UGC 03308&  05:26:01.8& 08:57:25& 9,836 -- 7,198&   1.2&6.5& A& ...\\
UGC 03585&  06:53:33.9& 27:18:32& 11,854 -- 9,216&  1.4&6.5& A& ...\\
UGC 03710&  07:09:32.0& 28:39:39& 8,988 -- 6,350&  0.44&6.5& A& ...\\
UGC 03790&  07:18:30.3& 31:22:30& 4,757 -- 2,119&  0.69&6.5& A& ...\\
UGC 04109&  07:56:16.7& 11:39:44& 15,071 -- 12,433&0.40&6.5& A& ...\\
UGC 04131&  07:59:11.8& 31:48:29& 38,500 -- -548&  0.48& 26& A& a,b,d,f,g,i,k--o,q\\
UGC 04144&  07:59:27.4& 07:26:37& 38,500 -- -548&  0.32& 26& A& a,b,d--h,k--o,q\\
UGC 04288&  08:14:35.8& 19:21:07& 31,745 -- -970&  0.49& 26& A& o,p,q,s,t\\
UGC 04496&  08:36:36.3& 25:08:12& 31,745 -- -970&  0.58& 26& A& ... \\
UGC 04831&  09:11:35.5& 32:50:55& 5,360 -- 2,992&  0.42&6.5& A& ... \\
UGC 05009&  09:24:44.6& 20:01:45& 5,596 -- 2,958&  0.41&6.5& A& ... \\
UGC 05211&  09:44:39.0&-00:13:17& 7,836 -- 5,198&  0.51&6.5& A& ... \\
UGC 05361&  09:59:00.8& 25:12:08& 31,745 -- -970&  0.58& 26& A& h,o,p,q,s,t\\
UGC 05440&  10:05:36.2& 04:16:23& 31,745 -- -970&  0.42& 26& A& h,o,p,q,s,t \\
UGC 05583&  10:20:35.1& 25:23:01& 31,745 -- -970&  0.83&6.5& A& ... \\
UGC 05592&  10:21:32.7& 22:32:47& 31,745 -- -970&  0.88& 26& A& ... \\
UGC 05679&  10:28:52.8& 26:20:11& 7,807 -- 5,169&  0.52& 26& A& ... \\
UFC 05770&  10:28:52.8& 26:20:11& 7,807 -- 5,169&  0.52& 26& A& ... \\ 
UGC 05710&  10:31:27.2& 24:08:34& 31,745 -- -970&  0.73& 26& A& i,m,o,p,q,s,t\\
UGC 05743&  10:34:48.8& 25:32:40& 31,745 -- -970&   1.2& 26& A& ...\\
UGC 05769&  10:37:02.7& 20:25:54& 31,745 -- -970&  0.45& 26& A& ... \\
UGC 05785&  10:38:26.0& 30:08:41& 7,684 -- 5,046&  0.56&6.5& A& ...\\
UGC 05790&  10:39:04.7& 04:38:51& 31,745 -- -970&  0.48& 26& A& j,m,o,p,q,s,t\\
UGC 05801&  10:39:52.7& 21:50:42& 31,745 -- -970&  0.49& 26& A& j,m,o,p,q,s\\
UGC 05828&  10:42:30.5& 15:45:52& 31,745 -- -970&  0.78& 26& A& j,l\\
\obc N02-2& 10:42:33.2& 24:41:35& 31,745 -- -970&  0.47& 26& A& j,m,o,p,q,t\\
\obc N02-3& 10:44:07.0& 24:42:32& 31,745 -- -970&  0.82& 26& A& j,m,o,p,q,t\\
UGC 05930&  10:49:34.9& 21:59:36& 31,745 -- -970&  0.86& 26& A& ... \\
UGC 06031&  10:55:00.6& 29:32:36& 31,745 -- -970&  0.97& 26& A& ...\\ 
UGC 06124&  11:03:39.5& 31:51:30& 31,745 -- -970&  0.64& 26& A& j,o,p,q,s \\
UGC 06243&  11:12:24.4& 31:24:41& 31,745 -- -970&  0.47& 26& A& j,m,o,p,q,s\\
UGC 06300&  11:17:16.9& 16:19:37& 31,745 -- -970&  0.49& 26& A& j,o,p,q,s \\
UGC 06401&  11:23:19.1& 13:37:47& 2,061 -- -577 &  0.62&6.5& A& ...\\
UGC 06524&  11:13:30.0& 23:10:37& 31,745 -- -970&  0.97& 26& A& t\\
UGC 06425&  11:24:44.8& 23:36:54& 8,064 -- 5,426&  0.74&6.5& A& ...\\ 
UGC 06425&  11:24:45  & 23:36:54& 9,278 -- 4,212 &  2.0& 17& N& ...\\ 
\end{tabular}
\end{table*}
\addtocounter{table}{-1}
\begin{table*}[p]
\centering
\caption{Observed Galaxies, {\it cont.}}
\begin{tabular}{llllcccl}
\hline\\[-10pt]
\hline\\
{\bf Galaxy}& {\bf RA} & {\bf Dec} & {\bf Search Range}
&{$\sigma$} & {\bf resolution} &{\bf Tel.} &{\bf RFI \dag}\\
& {$\left[J2000\right]$} & {$\left[J2000\right]$}&
{$\left[km s^{-1}\right]$} &{$\left[mJy\right]$} &{$\left[km\;s^{-1}\right]$} \\
\hline\\
UGC 06557&  11:35:07.5& 29:53:27& 31,745 -- -970&  0.54& 26& A& j,o,p,q,s,t\\
UGC 06659&  11:42:07.3& 32:32:21& 31,745 -- -970&  0.47& 26& A& j,l,m,o,p,q,s\\
UGC 06748&  11:46:24.3& 35:43:42& 31,745 -- -970&   1.3& 26& A& j,o,p,q,s\\
UGC 06842&  11:52:12.4& 34:47:39& 31,745 -- -970&   1.6& 26& A& l,q\\
UGC 06897&  11:55:36.4& 09:46:54& 7,843 -- 5,205&   1.1&6.5& A& ...\\
UGC 06913&  11:56:15.1& 17:01:44& 8,119 -- 5,481&   1.2&6.5& A& u\\
UGC 06947&  11:57:56.8& 22:11:26& 31,745 -- -970&  0.74& 26& A& s\\
\obc N10-2& 11:58:42.0& 20:34:43& 31,745 -- -970&  0.41& 26& A& ... \\
UGC 07084&  21:22:01.5& 07:08:43& 31,745 -- -970&   1.3& 26& A& u\\
UGC 07342&  12:18:30.8& 05:55:35& 31,745 -- -970&   1.3& 26& A& i,j,m,o,p,q,s\\
UGC 07369&  12:19:38.7& 29:53:00& 1,544 -- -1,094& 0.62&6.5& A& ... \\
UGC 07388&  12:20:13.8& 33:39:55& 31,745 -- -970&   1.4& 26& A& j,m,o,p,q,s,t \\
UGC 07425&  12:21:53.6& 15:38:46& 31,745 -- -970&   1.0& 26& A& m,o,p,q,t,u\\
UGC 07437&  12:22:19.5& 28:49:54& 31,745 -- -970&  0.89& 26& A& q\\
UGC 07438&  12:22:19.8& 30:03:48& 31,745 -- -970&  0.51& 26& A& j,m,o,p,q,s\\
UGC 07457&  12:23:09.8& 29:20:58& 31,745 -- -970&  0.85& 26& A& m,o,p,q,t,u\\
UGC 07598&  12:28:30.9& 32:32:51& 10,360 -- 7,722&  1.0&6.5& A& ...\\
UGC 07630&  12:29:43.8& 11:24:09& 31,745 -- -970&   1.3& 26& A& ...\\
UGC 07770&  12:36:17.0& 20:59:55& 31,745 -- -970&   1.0& 26& A& q\\
UGC 07928&  12:45:09.2& 23:02:21& 31,745 -- -970&  0.87& 26& A& s,u\\
UGC 07929&  12:45:20.8& 21:25:37& 31,745 -- -970&   1.2& 26& A& ...\\
UGC 07934&  12:45:38.7& 35:05:01& 31,745 -- -970&  0.76& 26& A& j,o,p,q,t\\
UGC 08081&  12:58:09.0& 14:51:32& 31,745 -- -970&  0.70& 26& A& ... \\
UGC 08171&  13:04:39.8& 18:25:29& 20,716 -- 18,078& 1.2&6.5& A& ... \\
UGC 08311&  13:13:50.7& 23:15:17& 4,771 -- 2,133&  0.69&6.5& A& v\\
UGC 08637&  13:39:36.6& 06:10:09& 31,745 -- -970&  0.83& 26& A& ...\\
UGC 08644&  13:40:01.5& 07:21:55& 38,500 -- -548&   2.3& 26& A& f,g,m,o,p,q,r,v\\
UGC 08762&  13:51:00.8& 24:05:27& 4,714 -- 2,076&  0.64&6.5& A& ... \\
UGC 08799&  13:53:19.9& 05:46:17& 31,745 -- -970&  0.90& 26& A& q\\
UGC 08802&  13:53:08.1& 35:42:41& 31,745 -- -970&  0.77& 26& A& i,m,o,q\\
UGC 08904&  13:58:51.1& 26:06:24& 31,745 -- -970&  0.51& 26& A& i,m,o,p,q,t\\
UGC 09008&  14:05:01.8& 11:00:42& 5,665 -- 3,027&  0.44&6.5& A& ...\\
UGC 09010&  14:05:20.4& 30:48:42& 8,650 -- 6,012&  0.63&6.5& A& ...\\
UGC 09238&  14:24:47.9& 35:16:29& 31,745 -- -970&  0.78& 26& A& i,m,o,p,q\\
UGC 09243&  14:25:33.2& 33:50:52& 4,636 -- 1,998&   1.3&6.5& A& ...\\
\obc A01-1& 14:27:07.4& 25:47:04& 31,745 -- -970&  0.80& 26& A& j,m,o,q,s\\
UGC 09513&  14:46:09.3& 13:01:45& 15,102 -- 12,464&0.83& 26& A& ...\\
UGC 09676&  15:03:30.6& 27:49:30& 4,203 -- 1,565&  0.83&6.5& A& ...\\
UGC 09680&  15:04:02.1& 18:38:56& 31,745 -- -970&  0.51& 26& A& i,m,o,p,q,t\\
UGC 09767&  15:12:48  & 07:26:02& 16,216 -- 11,150& 3.1& 17& N& ...\\
UGC 09770&  15:13:22.3& 25:11:50& 31,745 -- -970&  0.72& 26& A& i,o,p,q,s,t\\
UGC 10217&  16:07:40  & 22:20:31& 15,964 -- 10,898&0.44& 17& N& ... \\
UGC 10365&  16:24:05.9& 04:41:58& 31,745 -- -970&  0.64& 26& A& j,l,m,o,p,q,t\\
UGC 10377&  16:25:01.7& 23:04:12& 31,745 -- -970&  0.55& 26& A& j,l,m,o,p,q,t\\
UGC 10673&  17:03:08.6& 29:51:50& 31,745 -- -970&  0.56& 26& A& j,l,m,o,p,q,s\\
UGC 10674&  17:03:40  & 09:17:50& 13,030 -- 7,964&  1.2& 17& N& ... \\
UGC 11396&  19:03:49.5& 24:21:28& 38,500 -- -548&  0.50& 26& A& b,o,q\\
UGC 11569&  20:28:47.8& 10:38:02& 38,500 -- -548&  0.81&5.2& A& b,m,o,q\\
UGC 11625&  20:45:57  & 28:31:56& 17,103 -- 12,037& 2.4& 17& N& ...\\
UGC 11654&  20:58:12.1& 04:28:39& 38,500 -- -548&   1.4& 26& A& b,m,o,q,r\\
UGC 11694&  21:11:52  & 11:16:42& 7,598 -- 2,532&   1.1& 17& N& ...\\
UGC 11742&  21:26:26.3& 02:03:02& 38,500 -- -548&   1.9& 26& A& a--c,e--h,k--r,v\\
UGC 11840&  21:53:18  & 04:14:53& 10,519 -- 5,453&  5.2& 17& N& ...\\
UGC 12021&  22:24:12  & 06:00:13& 7,005 -- 1,939 &  1.4& 17& N& ...\\
UGC 12189&  22:48:06  & 03:55:41& 15,274 -- 10,208& 8.2& 17& N& ...\\
UGC 12359&  23:06:05  & 14:52:02& 13,236 -- 8,170 & 3.0& 17& N& ...\\
\end{tabular}
\end{table*}
\addtocounter{table}{-1}
\begin{table*}
\centering
\caption{Observed Galaxies, {\it cont.}}
\begin{tabular}{llllcccl}
\hline\\[-10pt]
\hline\\
{\bf Galaxy}& {\bf RA} & {\bf Dec} & {\bf Search Range}
&{$\sigma$} & {\bf resolution} &{\bf Tel.} &{\bf RFI \dag}\\
& {$\left[J2000\right]$} & {$\left[J2000\right]$}&
{$\left[km s^{-1}\right]$} &{$\left[mJy\right]$} &{$\left[km\;s^{-1}\right]$} \\
\hline\\
UGC 12424&  23:13:11  & 10:46:20& 13,155 -- 8,098 & 4.8& 17& N& ...\\
\obc\ P06-6&23:44:07.9& 09:13:40& 31,745 -- -970&   0.4&6.5& A& h,o,p,q,s\\
\hline\\
\multicolumn{8}{l}{
\dag RFI ($\pm$2.5 MHz, $\pm$260 \kms): 
a.~1240 MHz (38,077 \kms); b.~1246 MHz (36,810 \kms);}\\ 
\multicolumn{8}{l}{c.~1257 MHz (34,489 \kms); d.~1258 MHz (34,278 \kms); 
e.~1265 MHz (32,800 \kms);}\\ 
\multicolumn{8}{l}{f.~1267 MHz (32,378 \kms); 
g.~1276 MHz (30,478 \kms); h.~1278 MHz (30,056 \kms);}\\ 
\multicolumn{8}{l}{i.~1283 MHz (29,001 \kms); j.~1290 MHz (27,524 \kms); 
k.~1294 MHz (26,679 \kms);}\\ \multicolumn{8}{l}{l.~1303 MHz (24,780 \kms); 
m.~1312 MHz (22,880 \kms); n.~1320 MHz (21,119 \kms);}\\ 
\multicolumn{8}{l}{o.~1330 MHz (19,081 \kms); p.~1340 MHz (16,971 \kms); 
q.~1350 MHz (14,860 \kms);}\\
\multicolumn{8}{l}{ r.~1358 MHz (13,171 \kms); 
s.~1370 MHz (10,639 \kms); t.~1381 MHz (8,317 \kms) }\\
\hline\\[-10pt]
\hline\\
\end{tabular}
\end{table*}

\begin{table*}
\centering
\small
\caption{\ion{H}{1} Properties of Detected Galaxies \label{tab:hi_props}}
\begin{tabular}{lccccclcc}
\hline\\[-10pt]
\hline\\
{\bf Galaxy}& {$\rm \mathbf v_{HEL}$} & {$\rm \mathbf w_{20}$} &{$\rm \mathbf w_{50}$}
&{$\rm \mathbf \int flux  $} & {$\rm \mathbf log({{M_{HI}}\over{M_\odot}})$}&
{$\rm \mathbf v_{lit}$\dag} &{\bf Notes}& \\
& {$\left[ km\:s^{-1}\right]$} & {$\left[ km\:s^{-1}\right]$}&
{$\left[ km\:s^{-1}\right]$} &{$\left[ Jy\:km\:s^{-1}\right]$}& &
{$\left[ km\:s^{-1}\right]$} & {\ddag}
\\
\hline\\
UGC 0134 & 1684   & 157   & 141 & 2.44 & 8.5  & 764$^1$ &a\\
UGC 0189 & 7614   & 197   & 186 & 2.37 & 9.8  & 7618$^1$ &...\\
UGC 0667 & 6080   & 232   & 216 & 1.86 & 9.5  & 6086$^1$ &...\\
UGC 1122 & 21667  & 258   & 241 & 1.68 & 10.6 & 6086$^1$ &...\\
UGC 2299 & 10235  & 151   & 134 & 2.81 & 10.1 & 10253$^1$ &...\\
UGC 2580 & 9233   & 270   & 236 & 1.15 & 9.7  & ... &...\\
UGC 2588 & 10093  & 216   & 200 & 1.69 & 9.9  & ... &...\\
UGC 2641 & 7044   & 356   & 338 & 0.97 & 9.4  & ... &b\\
UGC 2671 & 7123   & 218   & 170 & 0.30 & 8.8  & 7127$^1$ &...& \\
UGC 2741 & 10985  & 64    & 48  & 1.59 & 9.9  & 11002$^1$ &...&\\
UGC 2796 & 9061   & 512   & 481 & 6.91 & 10.4 & 9076$^1$ &...&\\
UGC 03119& 7840 & 469& 430& 3.81& 10.0 & 7851$^1$ &...&\\
UGC 3225 & 11577  & 551   & 514 & 1.47 & 10.0 & ... &c\\
UGC 3308 & 8517   & 136   & 119 & 5.56 & 10.3 & ... &...\\
UGC 3585 & 10313  & 317   & 282 & 3.56 & 10.3 & 7685$^2$,10535$^3$ &...\\
UGC 3790 & 3459   & 251   & 217 & 2.91 & 9.2  & 3438$^1$ &c\\
UGC 4109 & 13696  & 347   & 326 & 2.44 & 10.3 & 13736$^2$,13752$^3$ &...\\
UGC 4131 & 17719  & 470   & 429 & 2.12 & 10.5 & ... &...\\
UGC 4144 & 9795   & 494   & 458 & 1.72 & 9.9  & ... &...\\
UGC 4288 & 30223  & 558   & 520 & 2.54 & 11.0 & ... &...\\
UGC 4496 & 8681   & 523   & 494 & 1.09 & 9.6  & 8666$^4$ &d\\
UGC 4831 & 4318   & 163   & 146 & 3.41 & 9.5  & 4311$^2$ &...\\
UGC 5009 & 4273   & 213   & 180 & 3.01 & 9.4  & 4277$^2$ &...\\
UGC 5211 & 6490   & 314   & 297 & 4.20 & 9.9  & 6517$^2$ &...\\
UGC 5440 & 18932  & 531   & 504 & 3.38 & 10.8 & ... &...\\
UGC 5592 & 7254   & 296   & 280 & 3.05 & 9.9  & ... &...\\
UGC 5679 & 6496   & 228   & 209 & 2.43 & 9.7  &  6488$^2$ &...\\
UGC 5743 & 5159   & 278   & 256 & 2.90 & 9.6  & ... &...\\
UGC 5769 & 12973  & 317   & 160 & 2.05 & 10.2 & ... &e\\
UGC 5770 & 12628  & 285   & 255 & 0.95 & 9.9  & 12638$^2$ &e\\
UGC 5785 & 6366   & 281   & 267 & 3.75 & 9.9  & 6365$^2$ &...\\
UGC 5801 & 16632  & 456   & 438 & 1.48 & 10.3 &  ... &...\\
UGC 5828-01 &15107& 552   & 539 & 1.65 & 10.3 & 14989$^2$&...\\
UGC 5828-02 &14617& 258   & 217 & 1.60 & 10.2 & 14608$^2$&...\\
UGC 5930 & 13222  & 265   & 228 & 2.03 & 10.2 & ... &...\\
UGC 6031 & 14504  & 560   & 525 & 4.69 & 10.7 & 14470$^1$ &...\\
UGC 6124 & 13970  & 613   & 583 & 2.19 & 10.3 &  ... &...\\
UGC 6243 & 12380  & 254   & 218 & 0.76 & 9.7  & ... &...\\
UGC 6300 & 1070   & 96    & 77  & 0.72 & 7.6  & ... &...\\
UGC 6401 & 888    & 69    & 53  & 0.35 & 7.1  & 742$^2$ &...\\
UGC 6425 & 6753   & 287   & 274 & 3.28 & 9.9  & 6745$^2$ &...\\
UGC 6524 & 6214   & 270   & 238 & 3.07 & 9.8  & ... &...\\
UGC 6525-01& 6061 & 126   & 103 & 1.21 & 9.3  & 6055$^5$,6079$^2$&...\\
UGC 6525-02& 6816 & 201   & 154 & 1.26 & 9.4  & ...& ...\\
UGC 6557 & 13880  & 350   & 330 & 0.83 & 9.9  & ... &...\\
UGC 6748 & 10775  & 282   & 250 & 2.56 & 10.2 & ... &...\\
UGC 6842 & 18807  & 433   & 414 & 1.65 & 10.4 & ... &...\\
UGC 6897 & 6525   & 255   & 235 & 4.57 & 10.0 &  6524$^1$ &...\\
UGC 6913 & 6795   & 468   & 446 & 4.33 & 10.0 & 6800$^2$ &...\\
UGC 6947 & 9253   & 400   & 375 & 3.01 & 10.1 & ... &f\\
UGC 7388 & 6461   & 471   & 439 & 3.05 & 9.8  & ... &...\\
UGC 7437 & 19597  & 539   & 499 & 2.13 & 10.6 & 19669$^1$ &g\\
UGC 7438 & 691    & 109   & 95  & 0.30 & 6.8  & ... &h\\
UGC 7598 & 9054   & 192   & 166 & 5.75 & 10.4 & 9041$^1$ &...\\
UGC 7770 & 8308   & 444   & 424 & 6.42 & 10.3 & ... &...\\
UGC 7934 & 9682   & 449   & 430 & 3.65 & 10.2 & ... &...\\
UGC 8171 & 19513  & 152   & 119 & 2.85 & 10.7 & 19397$^1$ &...\\
\end{tabular}
\end{table*}
\addtocounter{table}{-1}
\begin{table*}
\centering
\small
\caption{\ion{H}{1} Properties of Detected Galaxies {\it cont.}}
\begin{tabular}{lccccclcc}
\hline\\[-10pt]
\hline\\
{\bf Galaxy}& {$\rm \mathbf v_{HEL}$} & {$\rm \mathbf w_{20}$} &{$\rm \mathbf w_{50}$}
&{$\rm \mathbf \int flux  $} & {$\rm \mathbf log({{M_{HI}}\over{M_\odot}})$}&
{$\rm \mathbf v_{lit}$\dag} &{\bf Notes}& \\
& {$\left[ km\:s^{-1}\right]$} & {$\left[ km\:s^{-1}\right]$}&
{$\left[ km\:s^{-1}\right]$} &{$\left[ Jy\:km\:s^{-1}\right]$}& &
{$\left[ km\:s^{-1}\right]$} & {\ddag}
\\
\hline\\
UGC 8311 & 3493   & 266   & 202 & 2.85 & 9.2  &  3452$^1$ &...\\
UGC 8637 & 6958   & 147   & 129 & 4.39 & 10.0 & 6970$^2$ &i\\
UGC 8644 & 6983   & 152   & 142 & 1.05 & 9.4  & ... &...\\
UGC 8762 & 3407   & 181   & 166 & 5.71 & 9.5  & 3395$^1$ &...\\
UGC 8802 & 12342  & 417   & 396 & 3.80 & 10.4 & ... &...\\
UGC 8904 & 9773   & 300   & 278 & 4.63 & 10.3 & ... &...\\
UGC 9008 & 5348   & 162   & 147 & 2.67 & 9.6  & 4346$^2$ &...\\
UGC 9010 & 7350   & 206   & 188 & 2.15 & 9.7  & 7331$^1$ &j\\
UGC 9238 & 3348   & 220   & 201 & 0.95 & 8.7  & 3373$^2$ &k\\
UGC 9243 & 3339   & 215   & 201 & 4.25 & 9.4  & 3317$^1$ &...\\
UGC 9513 & 13926  & 242   & 213 & 1.34 & 10.1 & 13783$^2$ &l\\
UGC 9676 & 2887   & 176   & 175 & 3.78 & 9.2  & 2884$^1$ &...\\
UGC 9680 & 14572  & 549   & 504 & 1.75 & 10.3 & ... &...\\
UGC 10674-1& 10384 & 156  & 136 & 0.78 & 9.6  & 10497$^2$& m\\
UGC 10674-2& 10745 & 111  & 61  & 1.03 & 9.8  & 10497$^2$ & m\\
UGC 11396 & 4441  & 297   & 281 & 2.48 & 9.4  & ... &...\\
UGC 11569 & 4444  & 222   & 204 & 1.54 & 9.2  & ... &...\\
UGC 11694 & 4995  & -10   & 502 & 1.90 & 9.4  & 5065$^2$,5127$^3$ &n\\
UGC 11742 & 14559 & 188   & 163 & 0.90 & 10.0 & ... &...\\
UGC 12021 & 4472  & 275   & 259 & 6.74 & 9.8  & 4472$^6$,4502$^2$ &...\\
\obc\ A1-1 & 13891 & 262  & 242 & 1.16 & 10.0 & ... &...\\
\obc\ N2-2 & 13118 & 358  & 341 & 1.21 & 10.0 & ... &...\\
\obc\ N2-3 & 13105 & 360  & 345 & 1.19 & 10.0 & ... &...\\
\obc\ N10-2 & 20680 & 440 & 416 & 1.12 & 10.4 & ... &...\\
\\
\hline
\multicolumn{8}{l}{\dag Velocity Measurements:}\\ 
\multicolumn{8}{l}{$^1$Velocity is taken from NED or from Bottinelli et al. (\cite{bottinelli90}).}\\
\multicolumn{8}{l}{For these galaxies, no information other than the velocity is available from this catalog.}\\
\multicolumn{8}{l}{$^2$Optical velocity is from the {\it Updated Zwicky Catalog} (Falco et al. \cite{falco99}).}\\
\multicolumn{8}{l}{$^3$Optical velocity is from the CfA redshift survey (Markze, Huchra, 
\& Geller \cite{markze96}; Huchra, Vogeley,}\\
\multicolumn{8}{l}{\hskip 0.1in \& Geller \cite{huchra99}}\\
\multicolumn{8}{l}{$^4$21-cm velocity is from Giovanelli, Avera, \& Karachentsev (\cite{giov97a})}\\
\multicolumn{8}{l}{$^5$21-cm velocity is from Giovanelli et al. (\cite{giov97b})}\\
\multicolumn{8}{l}{$^6$H-$\alpha$ velocity from Mathewson, \& Ford (\cite{mathewson96})}\\
\multicolumn{8}{l}{\ddag Galaxy notes are given in Appendix A.}\\
\hline\\[-10pt]
\hline\\
\end{tabular}
\end{table*}

%
\appendix
\section{Notes on Individual Galaxies}
\begin{description}
\item[a. UGC 00134:] This detection is likely UGC 00132, which at v$_{HEL}$ = 1666 \kms\ and \am{4}{0} away
is within the Nan\c{c}ay beam.
\item[b. UGC 02641:] ARK 099, at v$_{HEL}$ = 7080 \kms\ and \am{1}{4} away, is within the Arecibo beam. Both
UGC 02641 and ARK 099 are of late (S?) morphological type and neither have published \HI\
spectra. It is likely that both galaxies contribute to the observed \HI\
spectra.
\item[c. UGC 03225, UGC 3790:] These objects are listed in NED as galaxy pairs.
\item[d. UGC 04496:] NRGb 004.06, at v$_{HEL}$ = 8730 \kms\ and \am{0}{1} away, is within the Arecibo beam.
It is likely that both galaxies lie within the observed \HI\ spectra.
\item[e. UGC 05769, UGC 05770] Both of these galaxies  lie within the observed \HI\ spectra.
\item[f. UGC 06947:] CGCG 127-113, at v$_{HEL}$ = 9106 \kms\ and \am{4}{2} away, is still within
the first Arecibo side lobe.  However, CGCG 127-113 is classified as an E0 galaxy
and so is unlikely to contribute much (if any) gas to the \HI\ spectra measured for UGC 06947.
\item[g. UGC 07437:] This object is  a galaxy pair with one galaxy at v$_{HEL}$ = 19669 \kms\ and the
other at 19319 \kms.  It is doubtful that the smaller member of the pair
(UGC 07437-01) is contributing significantly to the observed spectra, though, as
it is classified as an E galaxy while UGC 07437 is an Sbc galaxy.
\item[h. UGC 07438:] NGC 4308, at v$_{HEL}$ = 624 \kms\ and \am{5}{1} away, could lie within the
first side lobe of the Arecibo beam.  NGC 4308 is classified as an E galaxy, though, and
so is likely not contributing significantly to the total \HI\ gas found.
\item[i. UGC 08637:] UGC 08637 is listed as a galaxy pair within NED.  However, no velocity is given for its
companion and no evidence is visible within UGC 08637's spectra for contamination from
another galaxy.
\item[j. UGC 09010:] UGC 09010 has two companions which could be contaminating the observed
spectra -- UGC 09012 at v$_{HEL}$=7483 \kms\ and \am{3}{2} away and
MCG +05-33-050 at v$_{HEL}$ = 7628 \kms\ and \am{4}{1} away.  Both companion galaxies
fall within the Arecibo beam and first side lobe.  MCG +05-33-050 is classified as a
S0 galaxy and is not likely contributing significantly to the observed \HI\ flux.
UGC 09012, though, is a large, bright, spiral galaxy which probably has considerably
contaminated the observed flux.
\item[k. UGC 09238:] This galaxy has a companion galaxy, UGC 09235 at v$_{HEL}$ = 2973 \kms\ and
only \am{1}{9} away, well within the Arecibo beam.  It is likely that UGC 09235
is the reason for the apparently lopsided spectra seen for UGC 09238.
\item[l. UGC 09513:] UGC 09515 lies at 14011 \kms\ and \am{3}{0} from UGC 09513.  UGC 09515 is
listed as an S? galaxy, and could contain enough \HI\ mass to considerably contaminate
the \HI\ flux found for UGC 09513.
\item[m. UGC 10674:] is not listed as either being a galaxy pair or having a companion
within 50\arcmin.  However, the spectra clearly shows two separate profiles.
\item[n. UGC 11696:] UGC 11694 is part of a galaxy group which also contains UGC 11696
(v$_{HEL}$ = 5018 \kms, distance = \am{6}{3}, UGC 11700 (v$_{HEL}$
= 5036 \kms, distance = \am{11}{3}), CGCG 426-015 (v$_{HEL}$ = 5031 \kms,
distance=\am{13}{4}), and UGC 11697 (v$_{HEL}$ = 5086, distance = \am{23}{4}).
As this galaxy was observed with the Nan\c{c}ay telescope, it is likely to be contaminated
by one or more of UGC 11694's companions.
\end{description}
\end{document}